\documentclass[fleqn,usenatbib]{mnras}

\usepackage[T1]{fontenc}

\DeclareRobustCommand{\VAN}[3]{#2}
\let\VANthebibliography\thebibliography
\def\thebibliography{\DeclareRobustCommand{\VAN}[3]{##3}\VANthebibliography}

\usepackage{graphicx}	
\usepackage{amsmath}	
\usepackage{amssymb}	

\usepackage{newtxtext,newtxmath}
\usepackage{multirow}
\usepackage{rotating}

\newcommand{\lya}{Ly$\alpha$ }

\defcitealias{2021MNRAS.508.1262M}{MM21}
\defcitealias{2020MNRAS.499.1640M}{MM20}

\title[Template for the memory of reionization in \lya forest]{Separating the memory of reionization from cosmology in the Ly$\alpha$ forest power spectrum at the post-reionization era
}

\author[Montero-Camacho, Liu \& Mao]{
Paulo Montero-Camacho,$^{1,2}$\thanks{E-mail: pmontero@pcl.ac.cn (PMC)}
Yuchen Liu$^{2,3,4}$ and Yi Mao$^{2}$\thanks{E-mail: ymao@tsinghua.edu.cn (YM)} 
\\
$^{1}$Department of Mathematics and Theory, Peng Cheng Laboratory, Shenzhen, Guangdong 518066, China\\
$^{2}$Department of Astronomy, Tsinghua University, Beijing 100084, China\\
$^{3}$Department of Physics and Astronomy, University College London, Gower Street, London WC1E 6BT, UK\\
$^{4}$Cavendish Astrophysics, University of Cambridge, Cambridge, CB3 0HE, UK}

\date{Accepted XXX. Received YYY; in original form ZZZ}

\pubyear{2022}

\begin{document}
\label{firstpage}
\pagerange{\pageref{firstpage}--\pageref{lastpage}}
\maketitle

\begin{abstract}
It has been recently shown that the astrophysics of reionization can be extracted from the Ly$\alpha$ forest power spectrum by marginalizing the memory of reionization over cosmological information. This impact of cosmic reionization on the Ly$\alpha$ forest power spectrum can survive cosmological time scales because cosmic reionization, which is inhomogeneous, and subsequent shocks from denser regions can heat the gas in low-density regions to $\sim 3\times10^4$~K and compress it to mean-density. Current approach of marginalization over the memory of reionization, however, is not only model-dependent, based on the assumption of a specific reionization model, but also computationally expensive. 
Here we propose a simple analytical template for the impact of cosmic reionization, thereby treating it as a broadband systematic to be marginalized over for Bayesian inference of cosmological information from the Ly$\alpha$ forest in a model-independent manner. This template performs remarkably well with an error of $\leq 6 \%$ at large scales $k \approx 0.19 \ \textup{Mpc}^{-1}$ where the effect of the memory of reionization is important, and reproduces the broadband effect of the memory of reionization in the Ly$\alpha$ forest correlation function, as well as the expected bias of cosmological parameters due to this systematic. The template can successfully recover the morphology of forecast errors in cosmological parameter space as expected when assuming a specific reionization model for marginalization purposes, with a slight overestimation of tens of per cent for the forecast errors on the cosmological parameters. We further propose a similar template for this systematic on the Ly$\alpha$ forest 1D power spectrum. 

\end{abstract}

\begin{keywords}
dark ages, reionization, first stars -- intergalactic medium 
\end{keywords}



\section{Introduction}
The main survey of the Dark Energy Spectroscopic Instrument \citep[DESI; ][]{2022arXiv220510939A} has already started. DESI will provide a plethora of exquisite data that will help nail down the redshift evolution of the Hubble parameter, the energy budget of the Universe and the dark energy equation of state, constrain the sum of the neutrino masses and inflationary parameters, and many other scientific goals (see \citealt{2016arXiv161100036D} for a detailed description of the primary science goals).  

Great observational power in \lya surveys brings both familiar and new challenges that need to be addressed to understand the precision of new \lya forest measurements. In addition to data reduction and instrumental systematics that can jeopardise the ability of \lya forest surveys to learn about the Universe, there are several important astrophysical systematics such as broad absorption lines \citep{2019ApJ...879...72G, 2021arXiv211109439E}, clustering of ultraviolet sources \citep{2014PhRvD..89h3010P,2014MNRAS.442..187G}, damped Lyman-$\alpha$ systems \citep[e.g. ][]{2012JCAP...07..028F, 2022arXiv220100827W}, metal contamination \citep[e.g.][]{2017A&A...603A..12B}, relativistic effects \citep{2016JCAP...02..051I}, streaming velocities between baryons and dark matter \citep{2018MNRAS.474.2173H}, and thermal relics from both \ion{He}{II} \citep{2020MNRAS.496.4372U} and \ion{H}{I} reionization \citep{2019MNRAS.487.1047M}. These astrophysical systematics have the potential to bias the Bayesian inference of cosmological parameters and to underestimate the error budget associated with a given measurement. Hence the objective for any \lya forest survey -- e.g. DESI -- is to be able to marginalize over the impact of these systematics or to develop a direct separation technique.

\begin{figure}
    \centering
    \includegraphics[width=\linewidth]{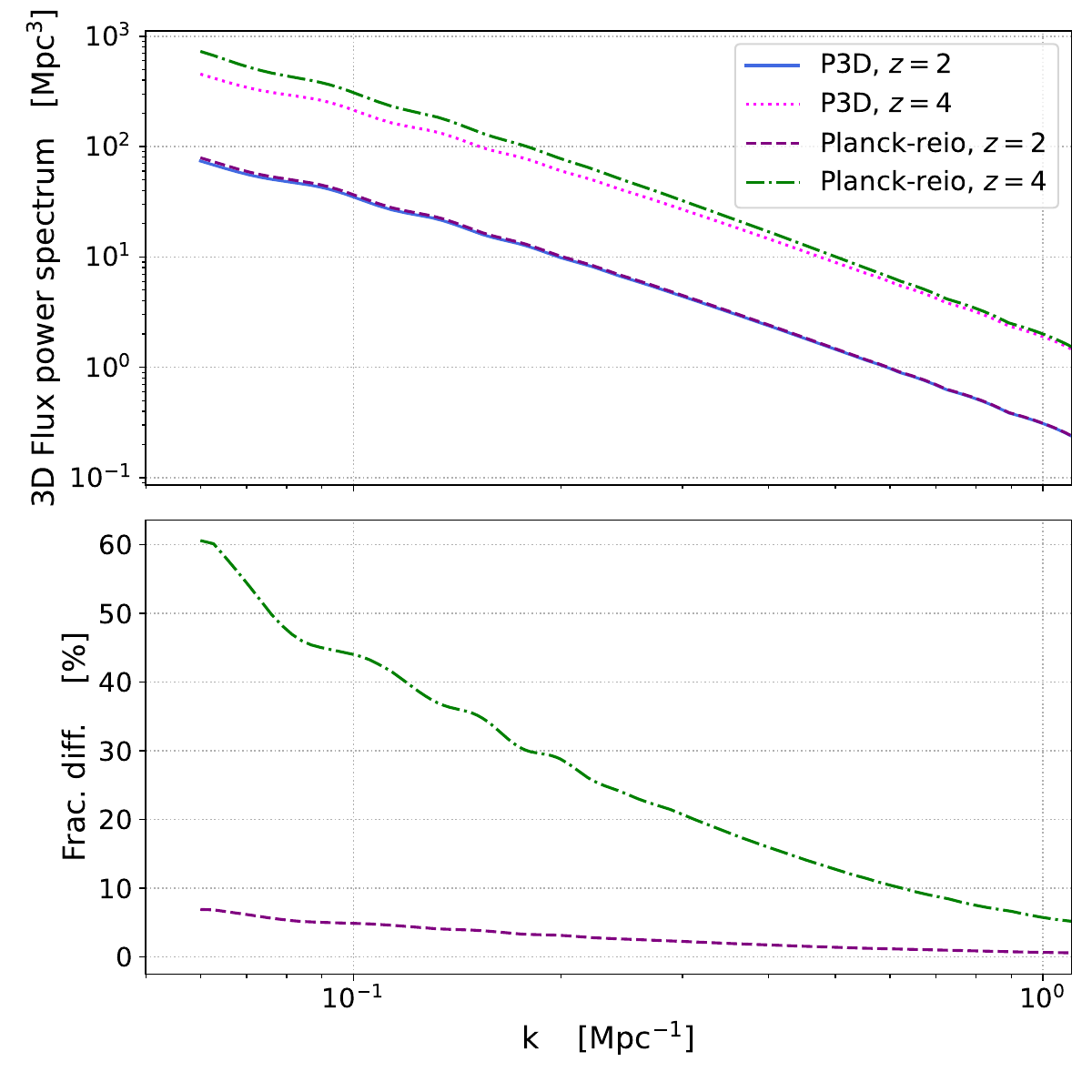}
    \caption{Impact of the memory of \ion{H}{I} reionization in the 3D \lya forest power spectrum (perpendicular to the line-of-sight, i.e. $\mu = 0$). (Top) Linear 3D flux power spectrum, i.e. first term in Eq.~(\ref{eq:PF3D}), at $z =4$ (magenta dotted line) and $z = 2$ (blue solid line) as a function of wavenumber, respectively. Also included is the total power spectrum including the enhancement due to the memory of reionization, i.e. the sum of the first and second term of Eq.~(\ref{eq:PF3D}), at $z = 4$ (green dash-dotted line) and $z = 2$ (purple dashed line), respectively. We choose a reionization scenario consistent with the midpoint of reionization measured by \citet{2020A&A...641A...6P}. (Bottom) Fractional difference of the total power spectrum with respect to the linear flux power spectrum at $z = 4$ (green dash-dotted line) and $z = 2$ (purple dashed line), respectively.}
    \label{fig:3D effect}
\end{figure}

\begin{figure}
    \centering
    \includegraphics[width=\linewidth]{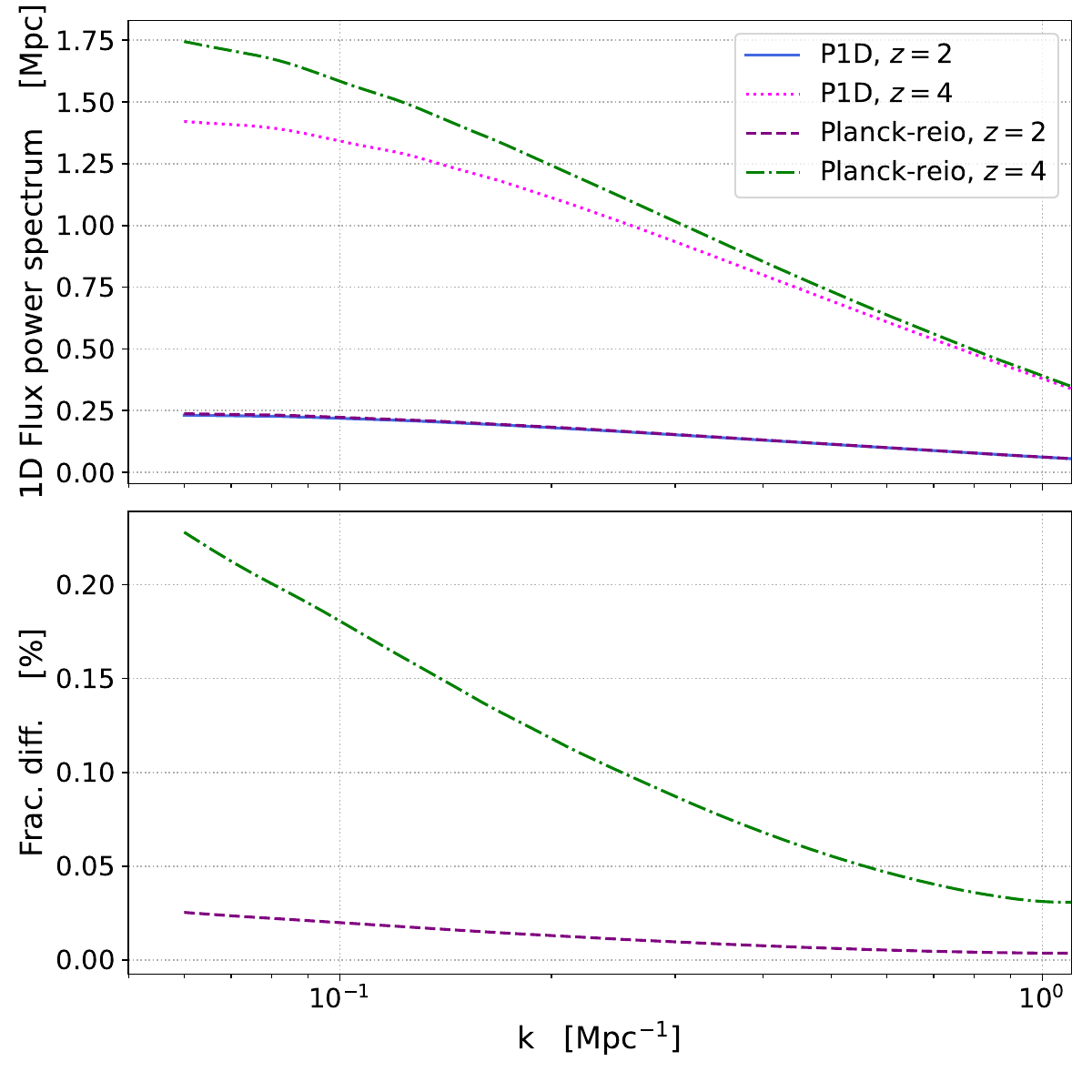}
    \caption{Same as Figure \ref{fig:3D effect} but for the memory of \ion{H}{I} reionization in the 1D \lya forest power spectrum. Here the linear 1D flux power spectrum (P1D) is given by the first term in Eq.~(\ref{eq:p1d}), and the total power spectrum, which includes the enhancement due to the memory of reionization, is the sum of the first and second term in Eq.~(\ref{eq:p1d}).}
    \label{fig:1D effect}
\end{figure}

Photoheating of the cold and neutral intergalactic medium (IGM) by ionization fronts 
\citep{2016ARA&A..54..313M} induces large-scale thermal fluctuations in the flux power spectrum that well extend to the post-reionization era \citep{2019MNRAS.486.4075O, 2019MNRAS.490.3177W, 2021arXiv210906897M}.\footnote{See also the results of \cite{2022ApJ...928..174M} where no enhancement at large scales was found; however, relics from cosmic reionization in the post-reionization era were still present. The reason for this disagreement is currently unknown.} These fluctuations are conventionally believed to dissipate by $z < 4$ to restore a rigorous temperature-density relation that rules over the temperature evolution of the IGM. However, \citet{2018MNRAS.474.2173H} found that the small-scale structure reionizes differently. In particular, underdense regions lead to a bi-modal temperature-density relation in the post-reionization IGM since low-density gas gets ultraviolet (UV) heated and shock-heated to high entropy. These shocks, which originate in the surrounding dense regions, also compress the underdense regions to the mean density, which is why this mode of the temperature-density relation is named the {\it high-entropy mean-density} (HEMD) phase. The HEMD gas can persist above the usual temperature-density relation until $z \sim 2$ \citep{2018MNRAS.474.2173H}. \citet{2019MNRAS.487.1047M} showed that the inclusion of the way the small-scale structure reionizes, i.e. accounting for the HEMD phase of the IGM temperature evolution, coupled with the patchy nature of cosmic reionization, produces a sizeable effect even at the relevant redshifts for current and near-term \lya forest surveys, i.e. $2.0 \leq z \leq 4.0$. We illustrate the effect of the memory of reionization in the 3D and 1D \lya forest power spectrum in Figures~\ref{fig:3D effect} and \ref{fig:1D effect}, respectively. Note that the memory of reionization is the combination of several physical effects: fluctuations in the ionizing radiation field, pressure smoothing caused by the violent heating during cosmic reionization that results in an increase in pressure that modulates the amount of baryonic structure, and the temperature fluctuations seeded by the inhomogeneous nature of reionization.

However, in order to fully capture the contribution of the way the small-scale structure reionizes, one must carefully track the small-scale structure below the Jeans mass prior to reionization ($\sim 10^6$ M$_\odot$). This feature makes traditional \lya simulations forbiddingly expensive. In fact, even the hybrid methodology proposed in \cite{2019MNRAS.487.1047M} is computationally expensive due to the required mass resolution to capture the role of the small-scale structure, and thus not suited for conventional Markov Chain Monte Carlo (MCMC) method. 
In addition, current approach of marginalization over the memory of reionization assumes a specific reionization model, to account for this effect explicitly (see, e.g. \citealt{2021MNRAS.508.1262M}; hereafter ``MM21'').   
Given the absence of a model-independent separation methodology and the required computational resources, marginalization over the effects of \ion{H}{I} reionization in the \lya forest would be theoretically and computationally challenging. 

Therefore, an analytical or semi-analytical prescription is needed to account for this astrophysical systematic. In this work, we propose an analytical template inspired by the Yukawa potential \citep{1935ppmsj...1....1Y}. We will demonstrate its ability to reproduce the memory of reionization, and test its forecast performance for the marginalization of the astrophysics of reionization with the broadband 3D \lya forest power spectrum using the Fisher matrix formalism. This template, therefore, provides a new working machinery that separates the memory of reionization from cosmology in the \lya forest power spectrum. In principle, it can be applied to the MCMC data analysis of real observations with reasonable amount of computational resources, in a model-independent manner. 

The rest of this paper is organized as follows. We introduce the Yukawa-like template for the memory of \ion{H}{I} reionization in the \lya forest power spectrum and examine its performance in \S\ref{sec:shape&perf}. 
We demonstrate the ability of our template to reproduce the memory of reionization in the \lya forest correlation function in \S\ref{sec:xi}, and, moreover, establish the degree of biasing expected due to this effect on cosmological parameters in \S\ref{sec:bias}. In \S\ref{sec:forecast}, we perform a proof-of-concept test for the marginalization of the astrophysics of \ion{H}{I} reionization using a Fisher matrix forecast for the broadband 3D \lya forest power spectrum using DESI specs (based on a description of the performance of the DESI spectrograph in \emph{g}-band and of the quasar luminosity function). 
We summarize our findings and discuss the directions for future work in \S\ref{sec:sum}. We leave some technical details to Appendix~\ref{app:priors} (testing the performance with different priors of reionization histories). Besides, in Appendix~\ref{app:p1d}, we apply a similar Yukawa-like template specialized for the memory of reionization in the 1D \lya forest power spectrum and describe its early implementation. 

Throughout this work we use the following fiducial cosmology: $\Omega_b = 0.0486$, $\Omega_m = 0.3088$, $\Omega_{\Lambda} = 0.6912$, and $h = 0.6774$.

\begin{figure*}
    \centering
    \includegraphics[width=\linewidth]{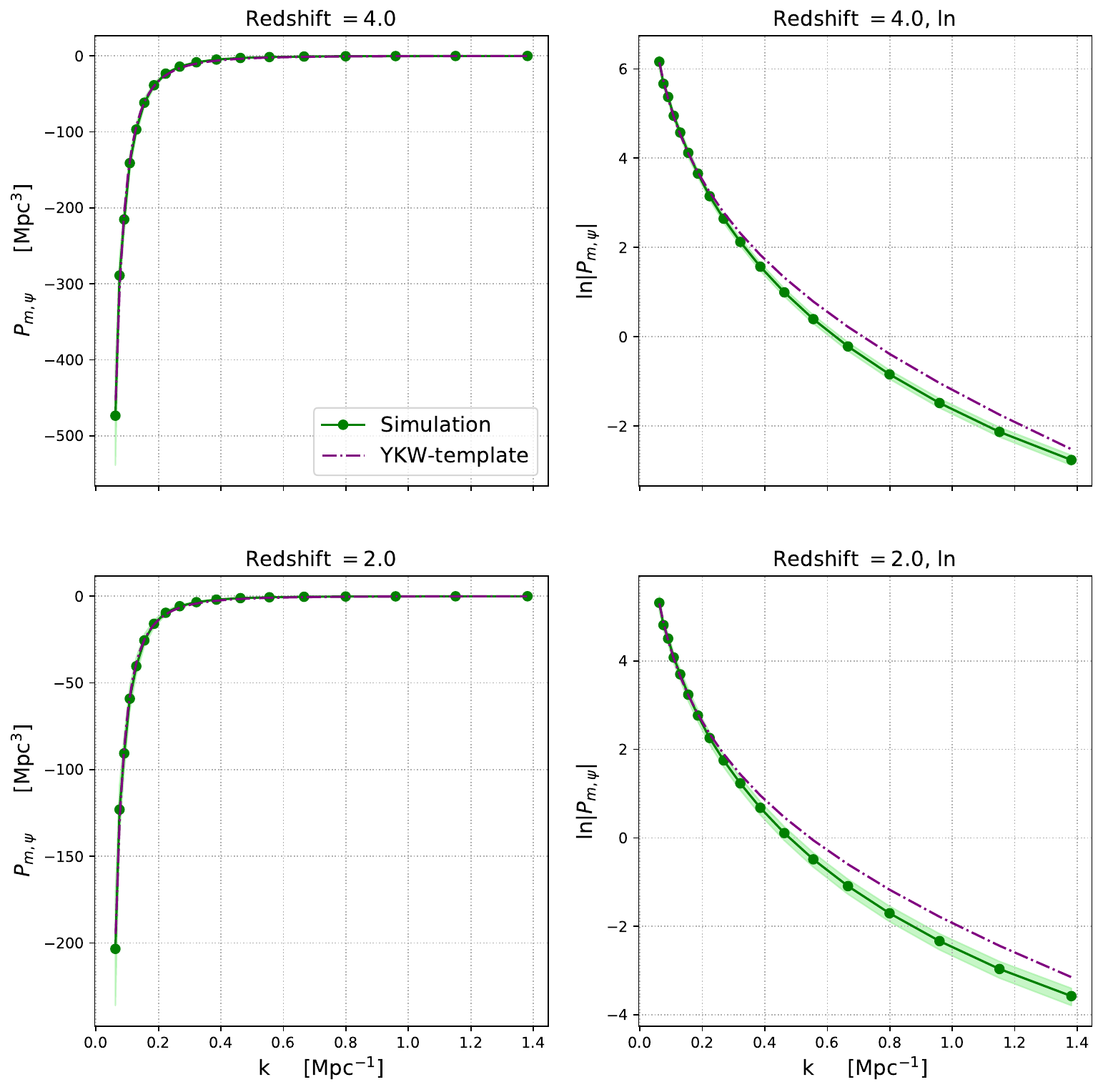}
    \caption{Performance of the Yukawa-like template for the memory of reionization imprinted in the \lya forest. Shown are the cross-power spectra of the matter and transparency of the IGM, $P_{m,\psi}(k,z_{\rm obs})$, at $z_{\rm obs} = 4.0 $ (top) and $z_{\rm obs}=2$ (bottom), in the linear scale (left) and in the natural logarithm (right). We show the results from the simulations of \citetalias{2020MNRAS.499.1640M} (green dots) as the fiducial model, with the light-green shaded area covering the error computed in \citetalias{2020MNRAS.499.1640M}. In comparison, we show the Yukawa-like template (purple dash-dotted line) that is the best-fit to the simulation results.}
    \label{fig:perf}
\end{figure*}

\section{Ansatz for the memory of \ion{H}{I} reionization}
\label{sec:shape&perf}
Following \citet{2019MNRAS.487.1047M}, the fluctuations on the transmitted flux field are given by
\begin{eqnarray}
    \label{eq:fluct}
    \delta_{\rm F} = b_{\rm F} (1 + \beta_{\rm F} \mu^2) \delta_{m} (k) + b_{\Gamma} \psi (k) \,
\end{eqnarray}
where $b_{\rm F}$ is the usual flux bias, $\beta_{\rm F}$ is the redshift-space distortion parameter, $\delta_m$ is the density contrast, $b_{\Gamma}$ is the radiation bias \citep[see, e.g.][]{2015JCAP...12..017A, 2018MNRAS.474.2173H}, which is necessary here to map from a fluctuation in optical depth to one in flux, and $\psi$ is a field that parametrizes the relative transparency of the IGM and it is formally a function of the redshift of observation, the local redshift of reionization, and the wavenumber. However, \cite{2019MNRAS.487.1047M} demonstrated that to compute the memory of reionization in the \lya forest only the overall modulation with respect to redshift of observation and local reionization redshift is needed -- see their Eq.~(4). Thus, we can define $\psi$ using the overall normalization factor utilized to match our simulations to the observed mean transmitted flux compared to that of a reference scenario as follows  
\begin{eqnarray}
    \label{eq:psi_ref}
    \psi(z_{\rm re},z_{\rm obs}) = \Delta \ln \tau_1 =  \ln \left[ \frac{\tau_1(z_{\rm re},z_{\rm obs})}{\tau_1(\bar{z}_{\rm re},z_{\rm obs}) }\right]\,
\end{eqnarray}
where $\tau_1$ is the optical depth that must be assigned in simulations to a patch of gas with mean density and temperature $T=10^4$ K in order for the mean transmitted flux to match observations, $z_{\rm re}$ stands for the local redshift of reionization while $\bar{z}_{\rm re}$ is a reference reionization scenario, here chosen to be 8. As commonly used in \lya studies, one varies the normalization of $\tau_1$, which is equivalent to varying the ionizing background, until the correct mean transmitted flux is obtained. Thus, any change in physics in our simulations, in our case the thermodynamic and hydrodynamic response of the IGM to the reionization process, is now summarized in the change of $\tau_1$. Here we use the observed flux from \cite{2007MNRAS.382.1657K}.

The transparency $\psi$ of the IGM is by construction a relative transparency with respect to a scenario with sudden reionization at redshift 8. As such,  $\psi(z_{\rm re}, z_{\rm obs}) > 0$ implies that a patch of gas that reionized at $z_{\rm re}$, and is observed at $z_{\rm obs}$, is more transparent than the fiducial scenario. Thus it requires a larger normalization to fit the observed transmitted flux. 

The 3D \lya forest power spectrum is then given by (\citetalias{2021MNRAS.508.1262M})
\begin{eqnarray}
    \label{eq:PF3D}
    P_{\rm F}(\boldsymbol{k},z) = b_{\rm F}^2 (1 + \beta_{\rm F} \,\mu^2)^2 P_{\rm L} D_{\rm NL} + 2 b_{\rm F} b_{\Gamma} (1 + \beta_{\rm F} \,\mu^2) P_{m,\psi} \, , 
\end{eqnarray}
where $P_{\rm L}(k,z)$ is the linear matter power spectrum. The first term corresponds to the conventional linear \lya flux power spectrum with the non-linear correction $D_{\rm NL}(k,\mu)$, which is given by \citep{2003ApJ...585...34M}
\begin{eqnarray}
    \label{eq:Dnl}
    \ln D_{\rm NL}(k, \mu) = \left(\frac{k}{k_{\rm NL}}\right)^{\alpha_{\rm NL}} - \left(\frac{k}{k_{\rm p}}\right)^{\alpha_{\rm p}} - \left(\frac{k_\parallel}{k_{\rm v} (k)}\right)^{\alpha_{\rm v}} \, .
\end{eqnarray}
The first term in Eq.~(\ref{eq:Dnl}) parametrizes the effect of non-linear growth on the flux power spectrum, the second one accounts for pressure smoothing, and the third term corresponds to the suppression due to peculiar velocities along the line of sight. We take the values for the different coefficients from \citet{2003ApJ...585...34M}.

We quantify the memory of reionization in the \lya forest -- which accounts for temperature fluctuations seeded by patchy reionization as well as fluctuations in the ionizing radiation field and pressure smoothing -- in terms of the cross-power spectrum of the matter and transparency of the IGM  \citep{2019MNRAS.487.1047M}
\begin{eqnarray}
\label{eq:psi}
P_{m,\psi}(k,z_{\rm obs}) = - \int {\rm d}z \frac{\partial \psi}{\partial z}(z,z_{\rm obs}) P_{m,x_{\rm HI}}(k,z)\frac{D_g(z_{\rm obs})}{D_g(z)}\, ,
\end{eqnarray}
where $\partial \psi/\partial z$ quantifies how the transparency of the IGM changes as a function of the redshift of reionization, $z$, for a given patch of the sky, as well as the redshift of observation $z_{\rm obs}$, which corresponds to the redshift of a \lya forest measurement, e.g. $2.0 \leq z_{\rm obs} \leq 4.0$ in the case of DESI. Note that $\psi$ enters Eq.~(\ref{eq:psi}) as an overall modulation factor and it does not depend on $k$. In fact, its spatial dependence ``has been absorbed'' by the cross-power spectrum of matter and neutral hydrogen field -- see the derivation of Eq. (4) in \cite{2019MNRAS.487.1047M}. This feature is what motivates the form of $\psi$ in Eq.~(\ref{eq:psi_ref}). 

Furthermore, $P_{m,x_{\rm HI}}$ is the cross-power between the matter and the neutral hydrogen fraction field, i.e. the correlation between matter distribution and ionized bubble spatial structure. Note that any change in the physics of reionization would be quantifiable \emph{locally} as a change of $\tau_1$, or analogously $\psi=\Delta \ln \tau_1$, and/or as a change on $P_{m,x_{\rm HI}}$ if it affects the large-scale structure. $D_g$ is the growth function. The integration in Eq.~(\ref{eq:psi}) covers the epoch of reionization.  We plot $P_{m,\psi}(k,z_{\rm obs})$ at representative redshifts of observation in Figure~\ref{fig:perf}. Although $P_{m,\psi} < 0$, the perturbation to the fluctuations of transmitted flux introduces a factor of flux bias in the cross-term; therefore, the effect on the flux power spectrum is an overall enhancement, particularly at large scales due to the coupling to the reionization bubble scale.

Based on the shape of $P_{m,\psi}(k,z_{\rm obs})$, we propose a Yukawa-like \citep{1935ppmsj...1....1Y} ansatz as given by
\begin{eqnarray}
\label{eq:yukawa}
P_{m,\psi} (k,z_{\rm obs}) = -\frac{A_{\rm re}}{(k/k_0)^{\beta_{\rm re}}} e^{-\alpha_{\rm re} (k/k_0)} \, ,    
\end{eqnarray}
where $A_{\rm re}$, $\alpha_{\rm re}$, and $\beta_{\rm re}$ are functions of redshift $z_{\rm obs}$. $A_{\rm re}$ is in units of Mpc$^3$, while $\alpha_{\rm re}$ and $\beta_{\rm re}$ are dimensionless. $k_0$ is a pivot wavenumber, and we choose $k_0 = 1\,{\rm Mpc}^{-1}$. Furthermore, these three free parameters also depend on the way reionization occurs. They are our proxy for how difficult is to form the parent galaxies that produce UV photons and how efficiently the UV photons escape into the intergalactic medium. 
    
Physically, $A_{\rm re}$ is the simplest parameter since it functions as a normalization factor, only involved with the overall amplitude of the cross-power spectrum. Analogous to the Yukawa potential, $\alpha_{\rm re}^{-1}$ corresponds to the effective range of the cross-power spectrum. We need to suppress $P_{m,\psi}$ at large $k$'s because the cross-power spectrum of matter and transmission of the IGM is really an integration of $P_{m,x_{\rm HI}}$ over the redshift of reionization, and the cross-power spectrum of matter and neutral hydrogen fraction couples to the reionization bubble scale. Naturally, we expect a stronger screening of the small scales at higher redshifts due to an overall stronger memory of reionization present at larger redshifts, given that there is not enough time for the IGM to recover the temperature-density relation \citep{1997MNRAS.292...27H}. 

We warn against the attempt to give a physical meaning of $\beta_{\rm re}$ since the analogous term in the Yukawa potential is designed by construction to match the massless propagator of electromagnetism, i.e. the photon that leads to the $r^{-1}$ Coulomb potential. Regardless, just as in the Yukawa potential, the denominator of Eq.~(\ref{eq:yukawa}) rules over the small-$k$ behaviour of the template by prevailing over the exponential decay at large scales.  

Note that the second term in Eq.~(\ref{eq:PF3D}) corresponds to the memory of reionization, where $P_{m,\psi}$ can now be given either by high-resolution simulations with Eq.~(\ref{eq:psi}) (e.g. \citealt{2020MNRAS.499.1640M}; hereafter ``MM20'') or by the template with Eq.~(\ref{eq:yukawa}).

The primary appeal of Eq.~(\ref{eq:yukawa}) is its robustness against specific prescriptions of how the IGM reionized, which often vary between different numerical codes that model the reionization process. In other words, if we allow for the ionization efficiency for UV photons to escape into the IGM to vary, either by its value or by functional form \citep[e.g. ][]{2019MNRAS.484..933P}, the template would still have the same shape (with, of course, different template parameters). Similarly, if we were to parametrize the difficulty of forming the star-forming galaxies responsible for ionizing the IGM with a different approach, e.g. constant minimum virial temperature versus halo mass cutoff, one would still have the same shape.

In Figure~\ref{fig:perf}, we show the performance of our template with respect to the fiducial model obtained from the hybrid simulations used in \citetalias{2020MNRAS.499.1640M}. The hybrid simulation combines the large-box simulation of inhomogeneous cosmic reionization with the seminumerical code {\sc 21cmFAST} \citep{2011MNRAS.411..955M,2019MNRAS.484..933P}, and the small-box structure formation simulation using a modified version of {\sc Gadget2} \citep{2005MNRAS.364.1105S} -- described and tested in detail in \cite{2018MNRAS.474.2173H} -- for accounting for the way small-scale structure reionizes.

Figure~\ref{fig:perf} shows that the template introduced here works quite well at large scales. This reflects the fact that the memory of reionization is strong at large scales. Specifically, the error at large scales is approximately 6\% at $k \approx 0.19$ Mpc$^{-1}$. However, this error increases to $\sim 40\%$ around $k = 0.5$ Mpc$^{-1}$, and is larger for smaller scales. 
The decrease in performance at larger $k$, although significant, will not be an issue to the implementation of this template for marginalization purposes, because of the small signal-strength of the memory of reionization at $k > 0.4$ Mpc$^{-1}$ (e.g. see the bottom panel of Figure~\ref{fig:3D effect}, or their Figure 5 of \citetalias{2020MNRAS.499.1640M}), as will be discussed in depth in \S\ref{sec:forecast}. 

We vary the astrophysical parameters, namely $T_{\rm min}$ (the minimum virial temperature of haloes that host ionizing sources) and $\zeta$ (the ionizing efficiency) -- as termed the ``bubble models'' in \citetalias{2020MNRAS.499.1640M}, because their variations directly affect the growth or evolution of the ionized bubbles. We find that the resulting cross-power spectrum of the matter and transmission of the IGM fits to our ansatz as well. This also provides an effective range for the parameters of our template for different representative reionization scenarios, as tabulated in Table~\ref{tab:temp_params}, that are all loosely consistent with observational constraints and upper limits (\citetalias{2020MNRAS.499.1640M}). Note that we do not include the variations of the parameter $R_{\rm mfp}$ (the maximum allowed size of the reionization bubbles) or the parameters that govern the astrophysics of the preheating of the IGM (the ``heating models'' in \citetalias{2020MNRAS.499.1640M}), because the cross-power spectrum of the matter and transmission of the IGM does not depend strongly on these parameters, as found in \citetalias{2020MNRAS.499.1640M}. 

\begin{table}
\centering
\caption{The range of template parameters in our ansatz for the memory of reionization in the \lya forest at both low ($z=2$) and high ($z=4$) redshifts. These values are obtained by fitting to the simulations used in \citetalias{2020MNRAS.499.1640M}, which span reionization scenarios with the midpoint of reionization between $6.97 \leq z_{\rm mid} \leq 8.28$. Likewise, the models span the values between $20 \leq \zeta \leq 30$ and $2 \times 10^4 \leq T_{\rm min} \leq 5 \times 10^4\,{\rm K}$ for the ionization efficiency and virial temperature, respectively.}
\label{tab:temp_params}
\begin{tabular}{ccc}
\hline\hline
Template parameter & $z = 2$ & $z = 4$ \\
\hline
$A_{\rm re}$ & [0.3047, 1.6322] & [0.9920, 4.0515] \\
$\alpha_{\rm re}$ & [0.7114, 2.8631] & [1.2326, 2.9322] \\
$\beta_{\rm re}$ & [1.8281, 2.2645] & [1.7942, 2.1752] \\
\hline\hline
\end{tabular}
\end{table}

\begin{figure*}
    \centering
    \includegraphics[width=\linewidth]{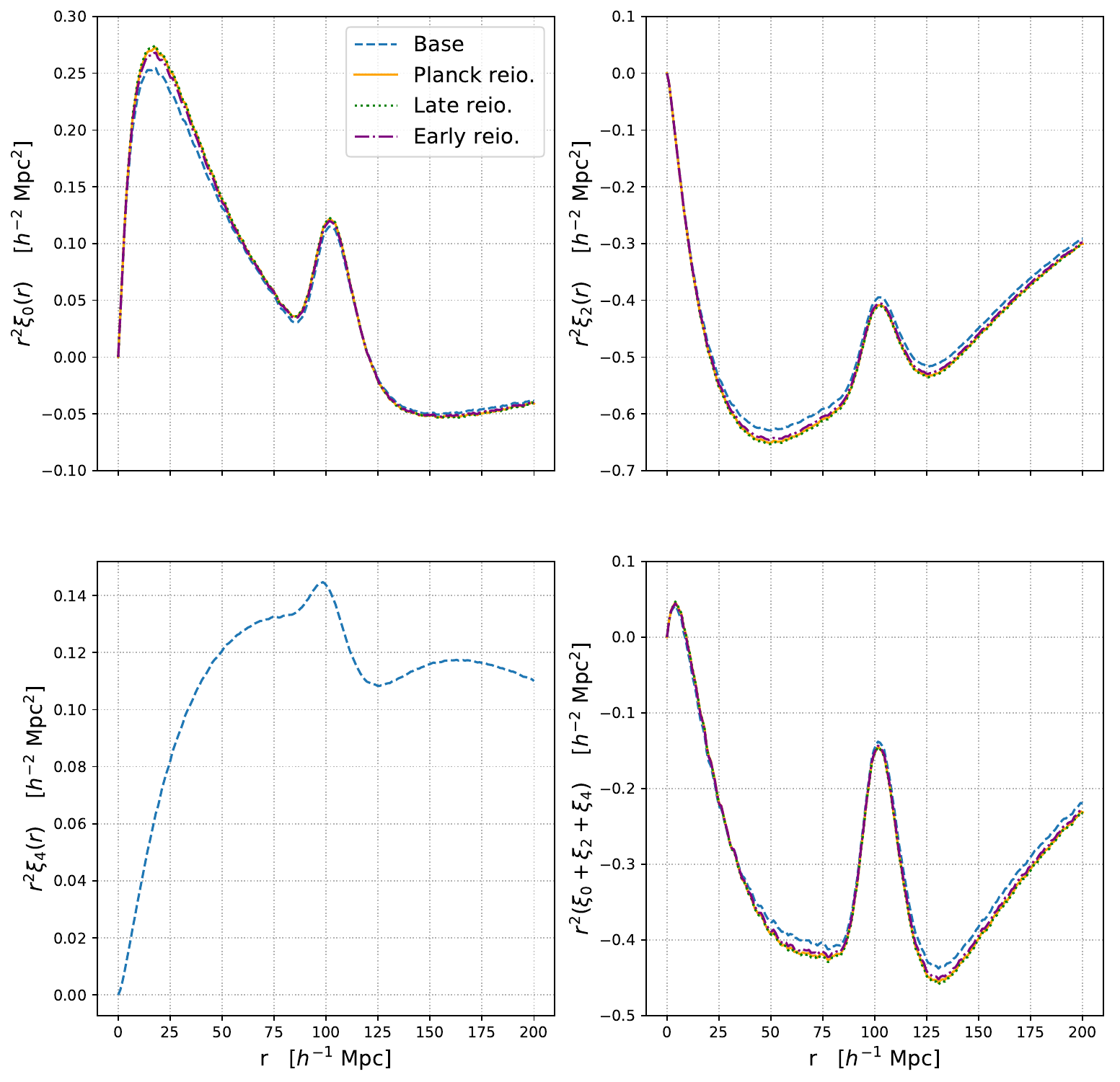}
    \caption{Multipoles of the \lya forest correlation function at $z = 2.25$. Shown are the monopole (top left), quadrupole (top right), and hexadecapole (bottom left). We also show the total correlation function (bottom right). We consider a ``base'' model (blue dashed line) which only contains the linear matter term (i.e.\ without the impact of reionization), and three models with memory of reionization: 
    the fiducial model of \citetalias{2020MNRAS.499.1640M} (``Planck reio'', orange solid line) in which the mid-point of reionization $z_{\rm mid} = 7.70$ is consistent with \citet{2020A&A...641A...6P}, the $\zeta_1$ model of \citetalias{2020MNRAS.499.1640M} (``Late reio'', green dotted line) and the $\zeta_2$ model of \citetalias{2020MNRAS.499.1640M} (``Early reio'', purple dash-dotted line) in which the mid-point of reionization $z_{\rm mid} = 7.21$/$8.08$ is consistent with 1$\sigma$ lower/upper value from Planck, respectively.
    The hexadecapole plot does not include any reionization models since it does not depend on the reionization term.}
    \label{fig:xi}
\end{figure*}

\begin{figure*}
    \centering
    \includegraphics[width=\linewidth]{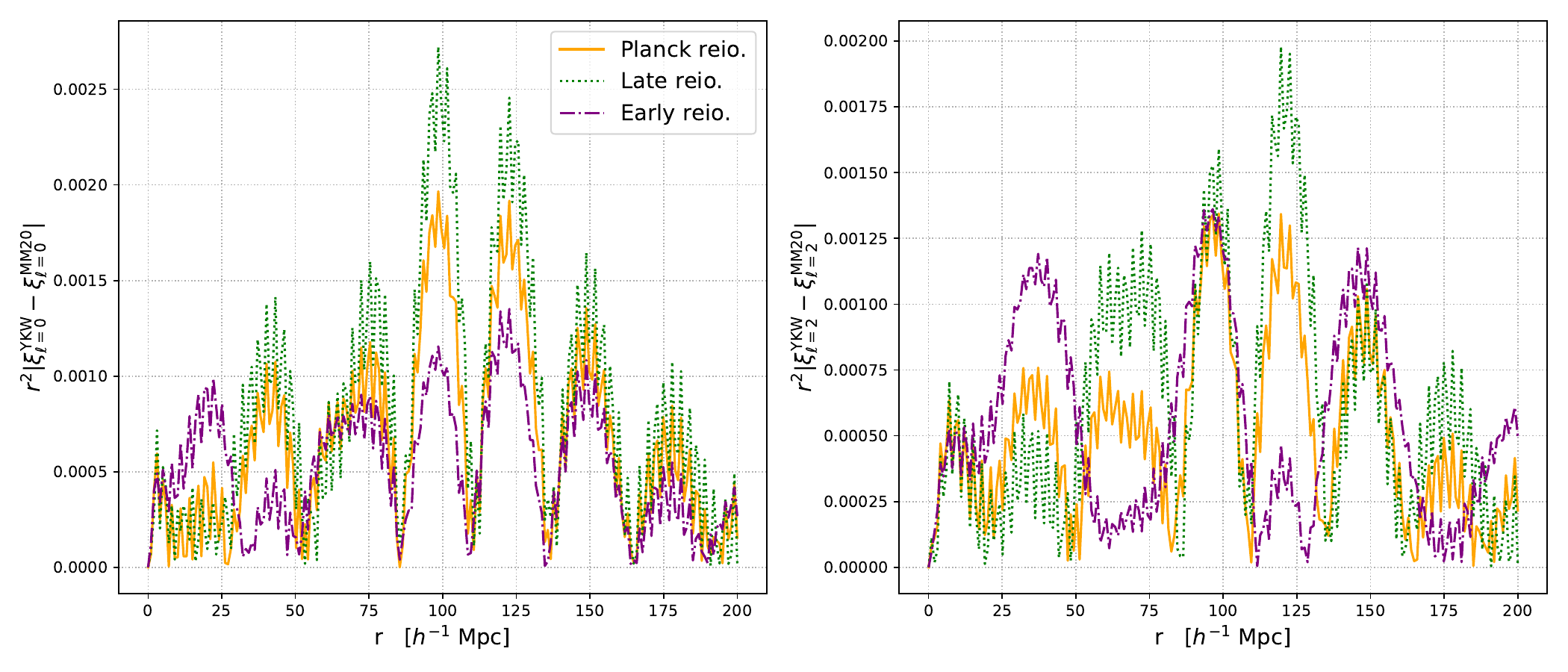}
    \caption{Performance of the Yukawa-like template in recovering the \lya forest correlation function. We compare the results that take into account the memory of reionization using the high-resolution simulations of \citetalias{2020MNRAS.499.1640M} and those using the Yukawa-like template, and show the absolute value of their difference in terms of the monopole (left) and quadrupole (right) of the \lya forest correlation function, respectively. We consider three reionization models as in Figure~\ref{fig:xi}.}
    \label{fig:xi_perf}
\end{figure*} 

\section{Impact of the Memory of reionization on correlation function and BAO feature}
\label{sec:xi}
Given the relative strength of the memory of reionization in the \lya forest at low redshifts \citep[e.g. ][]{2019MNRAS.487.1047M}, it is natural to ponder on the implications for the BAO feature in the \lya forest. Fortunately, reionization affects the \lya forest in a broadband fashion. Therefore, one should expect minimal contamination of the BAO peak or rather of its location. Here we justify this claim by computing the correlation function of the \lya forest using the simulations from \citetalias{2020MNRAS.499.1640M}. Furthermore, we demonstrate the capability of our proposed Yukawa-like template for reproduction of the memory of reionization in the correlation function. 

We follow the strategy in \cite{2014MNRAS.442..187G} to compute the multipoles of the correlation function, i.e.\  the non-linear corrections are neglected. The auto-correlation function of a cosmological tracer is given by \citep{2013JCAP...03..024K}
\begin{eqnarray}
    \label{eq:kirk1}
    \xi (r, \mu, z) = \sum_{\ell \, {\rm even}} L_{\ell} (\mu) \xi_{\ell} (r,z) \, , 
\end{eqnarray}
with $\xi_{\ell}$, the multipoles of the correlation function, given by
\begin{eqnarray}
    \xi_\ell (r,z) = \frac{i^{\ell}}{2\pi^2} \int_{0}^{\infty} {\rm d}k k^2 j_{\ell} (kr) P_\ell(k,z) \, , 
\end{eqnarray}
where $L_{\ell}$, $j_{\ell}$ are the Legendre polynomial and spherical Bessel function of degree $\ell$, respectively. Moreover, $P_{\ell}$ corresponds to the multipoles of the redshift space power spectrum of the \lya forest, which are given by
\begin{eqnarray}
    P_{0} & = & P_{\mu^0} + \frac{2}{3} P_{\mu^2} + \frac{1}{5} P_{\mu^4} \, , \\
    P_{2} & = & \frac{4}{3} P_{\mu^2} + \frac{4}{7} P_{\mu^4} \, , \\
    \label{eq:hexa}
    P_{4} & = & \frac{8}{35} P_{\mu^4} \, ,
\end{eqnarray}
where the moments are obtained from Eq.~(\ref{eq:PF3D}) -- ignoring the nonlinear correction, i.e. 
\begin{eqnarray}
    P_{\mu^0}(k,z) & = & b_{\rm F}^2 P_L (k,z) + 2 b_{\rm F} b_{\Gamma} P_{m,\psi}(k,z) \, , \\ 
    P_{\mu^2}(k,z) & = & b_{\rm F}^2 \beta_{\rm F} P_L(k,z) + b_{\rm F} b_{\Gamma} \beta_{\rm F} P_{m,\psi}(k,z) \, , \\
    P_{\mu^4}(k,z) & = & b_{\rm F}^2 \beta_{\rm F}^2 P_L (k,z) \, . 
\end{eqnarray}
We highlight that the hexadecapole of the power spectrum does not depend on reionization astrophysics, and thus extracting the multipoles of the \lya power spectrum will eventually become a promising novel way of mitigating the impact of inhomogeneous reionization in the \lya forest, when \lya forest observations become proficient enough at reliably extracting these multipoles. 

In Figure \ref{fig:xi}, we show the multipoles of the \lya forest auto-correlation function at $z = 2.25$, accounting for the memory of reionization using the high-resolution simulations in \citetalias{2020MNRAS.499.1640M} in three reionization models --- ``Planck reio''/``Late reio''/``Early reio'' in which the mid-point of reionization is consistent with the Planck/1$\sigma$ lower/upper value from Planck (\citealt{2020A&A...641A...6P}), respectively. For reference, we also include a ``base model'' without memory of reionization, i.e. only containing the first term in Eq.~(\ref{eq:PF3D}). There are two interesting features introduced by the inclusion of the memory of reionization in the \lya forest: (1) an enhancement of the amplitude (i.e.\ absolute value) of the correlation function, and (2) the robustness of the location of the BAO peak against this novel effect. The latter effect is not surprising due to the broadband nature of the memory of reionization in the \lya power spectrum. The former effect is particularly relevant at small distances ($r \sim 25 \ h^{-1}$ Mpc) for the monopole. In contrast, the quadrupole starts to show this trend at larger distances ($r > 25 \ h^{-1}$ Mpc). The reason for this discrepancy is the modulation by the spherical Bessel function of the first kind which in the case of the monopole quickly drops around this scale. On the other hand, $j_{2}$ (spherical Bessel function of $\ell =2$) starts at zero and rises around this distance. Besides, we highlight that the trend of these features has the expected hierarchical structure, i.e. late reionization leads to a stronger effect because the IGM has less time to relax into the usual temperature-density relation. 

A direct comparison of Figure \ref{fig:xi} with the results of \cite{2014MNRAS.442..187G,2014PhRvD..89h3010P} implies that the effect of the memory of reionization is smaller than the effect of the ionizing background on the \lya forest auto-correlation function at low redshifts. 
However, the situation is different at higher redshifts where the memory of reionization is stronger, because the HEMD gas is significantly not yet relaxed to the temperature-density relation.  

Finally, we plot the absolute error obtained by our proposed template when recovering the effect of reionization in the correlation function at $z = 2.25$ in Figure \ref{fig:xi_perf}. The absolute error is about an order of magnitude smaller than the effect of the memory of reionization for different reionization models at this redshift, so we deem this performance acceptable. 

\section{Impact of the Memory of reionization on cosmological parameter bias}
\label{sec:bias}
In this section, we proceed to investigate the robustness with which cosmological parameters can be constrained from \lya forest surveys under contamination from the memory of reionization. Throughout this section we focus on the scenario of a reionization history consistent with Planck's current constraints \citep{2020A&A...641A...6P} and consider only two cosmological parameters: $A_1= k_1^3 P_L(k_1) / (2\pi)^3$, the value of the dimensionless linear matter power spectrum evaluated at $k_1 \equiv 2 \pi\,h\,{\rm Mpc}^{-1}$, and $n_1$, the power-law index of the linear power spectrum evaluated at $k_1$. 

Our main objective here is to estimate the level of accuracy that our template is capable of for the parameter shifts $\Delta p_i$ of cosmological parameters ($A_1$ and $n_1$) compared to those inferred from the high-resolution simulations. The high-resolution simulations are taken from the suite of simulations used in \citetalias{2021MNRAS.508.1262M} for forecasting the ability of DESI to extract the astrophysics of reionization. In contrast to \citetalias{2020MNRAS.499.1640M}'s simulations, each simulation has only one realization, and hence for $k  \lessapprox 0.06 $ Mpc$^{-1}$ the estimates for the memory of reionization are not as precise, given the number of modes per wavenumber bin (see the wiggles at small $k$ in Figures 7 and 8 of \citetalias{2021MNRAS.508.1262M}). 

Following \citealt{2019MNRAS.485.5059U} (and references therein), the bias on a given parameter $p_i$ for a \lya forest survey that ignores the impact of inhomogeneous reionization is given by
\begin{eqnarray}
    \label{eq:shift}
    \Delta p_i(z) = \sum_j F_{ij}^{-1}(z) \Xi_j(z) \, ,
\end{eqnarray}
where $p_i = \{\bar{F}, A_1, n_1\}$\footnote{Note that we ignore the parameters that govern the temperature-relation, i.e.\ the amplitude of the power law $T_0$ and its tilt $\gamma$. We make this choice because the memory of reionization greatly disturb this power law \citep{2018MNRAS.474.2173H}.}, and $\bar{F} = \langle \exp{(- \tau)} \rangle$, the observed mean transmitted \lya flux. Here $F_{ij}$ is the Fisher matrix for a given redshift bin, which ignores the effect of memory of reionization, and the sum is over parameters. Moreover, $\Xi_j$ is given by
\begin{eqnarray}
    \label{eq:shift2}
    \Xi_j = \sum_{\rm bins} \frac{1}{\sigma_P^2} \frac{\partial P_{F}}{\partial p_j} \Delta P_F \, .
\end{eqnarray}
Here the sum over bins means over $k$ and $\mu$ bins, and does not include redshift bins since we focus on the impact at the lowest and highest redshift bins for clarity. $\Delta P_F$ parametrizes the new effect we want to consider, which in the present case is the memory of reionization in the 3D \lya forest flux power spectrum. Therefore, it is simply $\Delta P^{\rm 3D}_F = 2 b_F b_\Gamma (1 + \beta_F \mu^2) P_{m,\psi}$. Note that $P_{m,\psi}$ can be obtained either directly from the high-resolution simulations or via the Yukawa-like template.

The error on a given bin is a combination of cosmic variance, aliasing noise due to the sparse sampling of quasars, and spectrograph performance, 
\begin{eqnarray}
    \label{eq:error}
    \sigma^2_P = \frac{4 \pi^2 }{V_{\rm survey} k^2 \Delta k \Delta \mu} \left(P_F(\boldsymbol{k},z) + P_w^{2D}(z) P_F^{\rm 1D}(k_\parallel, z) + P_N^{\rm eff} \right)^2 \, ,
\end{eqnarray}
where $\Delta k$ and $\Delta \mu$ correspond to the width of the bins. The first term corresponds to Eq.~(\ref{eq:PF3D}) but without the memory of reionization, the second term is the aliasing term, which accounts for the sparse distribution of quasars, and the last terms is the expected spectrograph performance for DESI. See \cite{2014JCAP...05..023F}; \citetalias{2021MNRAS.508.1262M} for a detailed description of the noise terms. 

We obtain the Fisher matrix following the methodology described in \S 4 of \citetalias{2021MNRAS.508.1262M}, but turn off the contribution due to the memory of reionization and consider only the lowest and highest redshift bins, which correspond to $z_{\rm min} \approx 2.13$ and $z_{\rm max} \approx 3.94$, respectively.
\begin{eqnarray}
\label{eq:fisher}
F_{ij} = \sum_{k \geq k_{\rm min}}  \sum_{\rm \mu \, bins} \frac{1}{\sigma^2_P} \frac{\partial P_F}{\partial p_i} \frac{\partial P_F}{\partial p_j} \, ,
\end{eqnarray}
where we have also implemented a cutoff for minimum wavenumber $k_{\rm min} = 0.063 $~Mpc$^{-1}$ due to the small fraction of modes in the simulations used to obtain the parameters of the template for larger scales.

\begin{table}
    \centering
    \caption{Biases on the cosmological parameters $A_1$ (the amplitude of the linear matter power spectrum at $k_1$) and $n_1$ (power-law index of the linear power spectrum evaluated at $k_1$) introduced when neglecting the memory of reionization in the 3D \lya forest power spectrum. Shown are $\alpha_{1\sigma} = \Delta p_i^{\rm sims} / \sigma_i$ (the bias as a fraction of the forecast $1\sigma$ error), and ${\cal E} = | (\Delta p_i^{\rm YKW} - \Delta p_i^{\rm sims}) / \Delta p_i^{\rm sims}|$ (the percentage error on the performance of the template to reproduce the parameter shifts with respect to the simulations).}
    \begin{tabular}{cccc}
    \hline \hline
    {} & {} & $\alpha_{1\sigma}$ & ${\cal E}$ [\%] \\
    \hline
    \multirow{2}{*}{$z_{\rm min} \approx 2.13$} & $A_1$ & $-0.11$ & 1.85 \\
    {} & $n_1$ & $-1.16$ & 0.70  \\
    \hline
    \multirow{2}{*}{$z_{\rm max} \approx 3.94$} & $A_1$ & 0.25 & 3.37  \\ 
    {} & $n_1$ & 0.18 & 1.33  \\
    \hline \hline
    \end{tabular}
    \label{tab:p_shifts}
\end{table}

\begin{table}
    \centering
    \caption{Same as Table~\ref{tab:p_shifts} but for the 1D flux power spectrum.}
    \begin{tabular}{cccc}
    \hline \hline
    {} & {} & $\alpha_{1\sigma}$ & ${\cal E}$ [\%] \\
    \hline
    \multirow{2}{*}{$ z_{\rm min} = 2.2$} & $A_1$ & 0.02 & 4.17 \\
    {} & $n_1$ & $-0.21$ & 11.9 \\
    \hline
    \multirow{2}{*}{$z_{\rm max} = 4.0$} & $A_1$ & 0.49 & 2.11 \\ 
    {} & $n_1$ & $-0.96$ & 0.20 \\
    \hline \hline
    \end{tabular}
    \label{tab:p_shifts_1D}
\end{table}

In Table~\ref{tab:p_shifts}, we report the bias in the cosmological parameters, introduced when neglecting the memory of reionization in the 3D \lya forest power spectrum. Specifically, we compute $\alpha_{1\sigma} = \Delta p_i^{\rm sims} / \sigma_i$, the cosmological parameter bias estimated from high-resolution simulations of \citetalias{2021MNRAS.508.1262M} using Eqs.~(\ref{eq:shift}) --- (\ref{eq:error}) as a fraction of forecast $1\sigma$ errors. To test the performance of our template to reproduce the parameter shifts, we also compute the percentage error on the cosmological parameter bias with respect to the simulation results, ${\cal E} = | (\Delta p_i^{\rm YKW} - \Delta p_i^{\rm sims}) / \Delta p_i^{\rm sims}|$. 

At low redshift, neglecting the memory of reionization primarily affects the tilt of the matter power spectrum, where it introduces a bias larger than the $1\sigma$ forecasted error obtained from a Fisher matrix analysis that ignores this effect. The negative sign indicates that the parameter is expected to change to smaller values when neglecting the memory of reionization. There is a different trend at high redshift, $A_1$ is now slightly more impacted than $n_1$. Besides both shifts are positive. The change of the sign could be due to the larger effect of reionization at higher redshift and/or because of the loss of constraining power given by the wavenumber cutoff that we have introduced here since this feature is present using both the template and the high-resolution simulations. 

Table \ref{tab:p_shifts} showcases the success of the template (percentage difference $\leq 3.37$ \%) to account for the effects of ignoring the memory of reionization at both low and high redshifts. An important caveat here is the fact that we have assumed a known reionization history. In \S\ref{sec:forecast}, we change our focus to considering the impact of marginalization over astrophysics of reionization and discuss the limitations of our template. 

On the other hand, substantial medium \citep{2019JCAP...07..017C} and high resolution \citep{2019MNRAS.482.3458M, 2021AJ....161...45O} \lya spectra already exists. Furthermore, a plethora of upcoming data will be available in the near future \citep{2016arXiv161100036D, 2016sf2a.conf..259P}. Thus, there is a strong motivation to account for the memory of reionization in the easier-to-measure (because of the sparse sampling of quasars) --- one-dimensional \lya flux power spectrum, where one averages the three-dimensional flux over the perpendicular direction to the line of sight. In Table~\ref{tab:p_shifts_1D}, we tabulate our findings for the 1D version of the Yukawa-like template that is discussed in detail in Appendix~\ref{app:p1d}. 

For the 1D case, we obtain $\sigma_P$ by adding in quadrature the statistical and systematic errors tabulated in \citealt{2019JCAP...07..017C}, i.e.\ from the Extended Baryon Oscillation Spectroscopic Survey (eBOSS). In addition, we also use the wavenumber and redshift bins consistent with the eBOSS measurement. As a consequence, the minimum redshift bin is now $z_{\rm min} = 2.2$ and the maximum is $z_{\rm max} = 4.0$. Again, we see a stronger impact on the tilt than the amplitude at low redshift. However, the impact is severely diminished reaching only a shift of -0.21$\sigma$ due to the weaker effect of reionization in the one-dimensional flux power. In contrast, we see a (negative) shift of almost a sigma for $n_1$ and half a sigma for $A_1$ at high redshift. 
Interestingly, we no longer have a change of sign for the parameters at higher redshift. We argue that this is another sign of the impact of the cutoff applied in the 3D case. For the 1D case, there is no longer a cutoff since the power spectrum, particularly the term due to the memory of reionization, is significantly smoother at large scales because of the integration. 

Interestingly, the consequence of neglecting the memory of reionization in the 1D power spectrum could correct some known trends from \lya forest observations, such as (1) the preference for smaller values of $n_s$ (hence smaller values of $n_1$) than CMB predictions \citep[e.g. ][]{2015JCAP...11..011P,2020JCAP...04..038P}\footnote{Preliminary estimates have shown that ignoring reionization would lead to a negative shift for the running on $n_s$. Therefore, the inference of a negative running in eBOSS could also be likely explained by neglecting the memory of reionization. We leave a more accurate treatment of the running of the spectral index and its bias to future work.}, and (2) typical high-redshift measurements of the \lya forest on larger $\sigma_8$ values, which correspond to larger $A_1$, than low redshift measurements. These two trends in \lya forest observations could be simultaneously explained by the neglect of the memory of reionization. 

The 1D version of the template performs worse than the 3D template (percentage difference $\leq 11.9$ \%) at low redshift. Fortunately, this redshift bin also corresponds to small shifts in the cosmological parameters. At high redshift, where we observe the strongest bias, we see a significant boost in the performance of the template (percentage difference $\leq$ 2.11 \%). 

Hence, we deem the template successful enough for both 3D and 1D \lya forest --- we have tested this in a scenario where the reionization history is known by other means and not simultaneously fit from \lya spectra, but will further test it more generally by simultaneously marginalizing the template parameters (which reflect the memory of reionization) over cosmology in \S\ref{sec:forecast} for the 3D case and in Appendix \ref{app:p1d} with the first iteration of the 1D Yukawa-like template.

\section{Fisher forecast: template versus high-resolution simulations}
\label{sec:forecast}
Following \cite{2003ApJ...585...34M,2007PhRvD..76f3009M,2014JCAP...05..023F}; \citetalias{2021MNRAS.508.1262M}, we need three main ingredients to make a realistic forecast on the capability with which a \lya observation with the measurement of $P^{\rm 3D}_{\rm F}$ can constrain the underlying physics: (1) Quasar Luminosity Function (QLF), which quantifies how many \lya quasars we expect to have in a representative volume of the Universe \citep{2013A&A...551A..29P,2020RNAAS...4..179Y}, (2) spectrograph performance of DESI\footnote{There is no public description of the DESI spectrograph yet.}, and (3) a model for the observational signal as a function of the astrophysics of reionization, cosmology, and \lya astrophysics. 

\citetalias{2021MNRAS.508.1262M} presented a Fisher forecast for extracting the astrophysics of reionization from the \lya forest power spectrum, and showed how the thermal relics from \ion{H}{I} reionization affects the extraction of cosmological information from the broadband signal. This forecast was based on a suite of high-resolution, expensive simulations, which assumes a specific reionization model in order to account for the memory of reionization explicitly. Now, the analytical template we introduce in \S\ref{sec:shape&perf} provides an effective tool to separate the astrophysics of reionization from cosmology by marginalization of the template parameters over cosmological parameters in a phenomenological manner, so this new approach is computationally economic, and model-independent, i.e.\ not depending on the specific reionization model. 
The aim of this section is to establish how well this new approach can recover the change in forecast errors due to the thermal relics from reionization, i.e.\ how effective the marginalization over reionization astrophysics is with this mitigation strategy. 

Here, we no longer apply the cutoff for minimum $k$ used in Eq.~(\ref{eq:fisher}) in order to facilitate the comparison with the forecast results of  \citetalias{2021MNRAS.508.1262M}. This choice might introduce discrepancies between simulations and the template due to the low number of modes present in the simulations to compute $P_{m,\psi}$ at $k \lesssim 0.06 $ Mpc$^{-1}$. However, it will allow the Fisher matrix to capture the scales where cosmological information is best constrained. We emphasize that this is not a limitation of the Yukawa-like template and it could be overcome via larger box simulations than the ones used in both \citetalias{2020MNRAS.499.1640M} and \citetalias{2021MNRAS.508.1262M} to extract the template parameters. Alternatively, it would also suffice to employ a simple biasing procedure at larger scales where the memory of reionization follows the matter power spectrum up to a bias parameter. 

Furthermore, $P_F$ in Eq.~(\ref{eq:fisher}) is now given by Eq.~(\ref{eq:PF3D}), which includes the memory of reionization, and the vector of parameters to be forecasted is $p_i = \{\bar{F}, A_1, n_1, A_{\rm re}, \alpha_{\rm re}, \beta_{\rm re} \}$.  

\begin{table}
\centering
\caption{Fiducial values and variations for the \lya astrophysics and cosmological parameters of the forecast at $z_{\rm ref} = 2.25$. Here $\bar{F}$ is the mean transmitted flux, $A_1$ and $n_1$ correspond to the amplitude and tilt of the linear matter power spectrum at $k_1 \equiv 2 \pi\,h\,{\rm Mpc}^{-1}$, respectively. Note that the variation of a given parameter implies that the corresponding Fisher matrix element is computed with, e.g., $\bar{F}$ and $\bar{F} \pm \delta \bar{F}$.} For the template parameters, see Table \ref{tab:temp_params}.
\label{tab:params}
\begin{tabular}{ccc}
\hline\hline
Parameter $\boldsymbol{\theta}$ & Fiducial value & Variation $ \delta$\\
\hline
$\bar{F}$ & 0.844 & 0.05 \\
$A_1$ & 1.48 & 0.29 \\
$n_1$ & -3.19 & 0.10 \\
\hline\hline
\end{tabular}
\end{table}

In Table \ref{tab:params}, we summarize the fiducial values and variations on the non-reionization parameters of the forecast. We follow \citet{2003ApJ...585...34M} on modeling the base \lya forest, i.e. the term without relics from reionization, using their Table 1, and evolve with redshift by multiplying the power spectrum at the redshift pivot of $z_{\rm ref} = 2.25$ with $[(1+z)/(1 +z_{\rm ref})]^{3.55}$, based on Eq.~(14) of \citet{2013A&A...559A..85P}. For the astrophysics of reionization, i.e. the template parameters, we use the values tabulated in Table \ref{tab:temp_params} (and intermediate results at other redshifts not shown for brevity) to interpolate and compute the fiducial values and variations as a function of redshift bin. Models with late (early) reionization prefer large (small) values of $A_{\rm re}$ and $\alpha_{\rm re}$, and small (large) values of $\beta_{\rm re}$. These preferences are a consequence of the need for a stronger amplitude for late reionization scenarios, which then require a stronger reduction of the effective scale $(1/\alpha_{\rm re})$ to not overestimate the small-scale impact. Furthermore, late reionization requires an enhancement of $P_{m,\psi}$ at large scales, and thus a smaller $\beta_{\rm re}$. 

We check that our results do not depend strongly on the choice of variations for $A_{\rm re}$, $\alpha_{\rm re}$, and $\beta_{\rm re}$. For reference, decreasing our choice by a factor of two leads to a change in the forecast errors of $< 2 \%$. Similarly, the derivatives -- needed for the Fisher matrix calculation -- individually show deviations of the same order. Interestingly, the change was mainly present in $\beta_{\rm re}$ since the Fisher matrix constrains it the best among the template parameters. We attribute this feature to the role of $\beta_{\rm re}$ because modulating the large-scale behavior of the memory of reionization coincides with the scales where the thermal relics from reionization are significant.

\begin{figure}
    \centering
    \includegraphics[width=\linewidth]{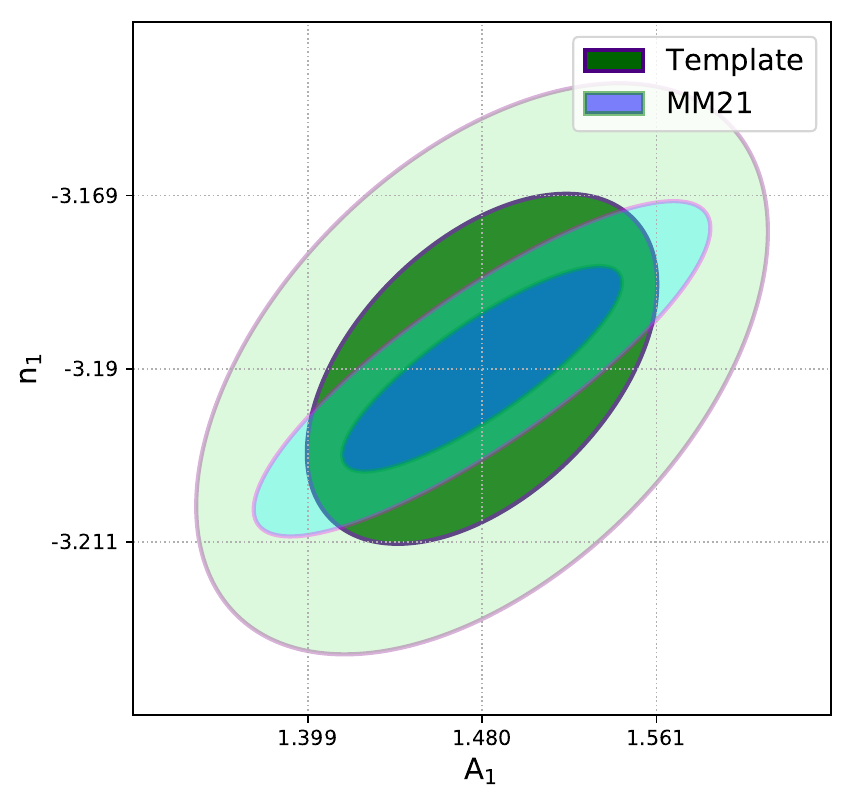}
    \caption{Forecast for extraction of cosmology from the broadband 3D \lya forest power spectrum with DESI by marginalizing over the astrophysics of reionization. Here, $A_1$ and $n_1$ are the amplitude and tilt of the dimensionless matter power spectrum evaluated at $k_1 \equiv 2\pi h $ Mpc$^{-1}$, respectively. We show the 1$\sigma$/2$\sigma$ contours obtained by \citetalias{2021MNRAS.508.1262M} using high-resolution simulations (blue/cyan ellipses), and the 1$\sigma$/2$\sigma$ contours obtained using the Yukawa-like template (green/light-green ellipses), respectively.}
    \label{fig:forecast}
\end{figure}

We show the ability of our template to extract cosmology from the 3D \lya forest power spectrum in Figure \ref{fig:forecast}, i.e. the contour for the amplitude of the linear matter power spectrum versus its tilt. The template can successfully recover the morphology of forecast errors in cosmological parameter space as expected in the high-resolution simulations of \citetalias{2021MNRAS.508.1262M}, with a slight overestimation of the errors. This decrease in performance is expected due to three reasons as follows.\footnote{Small discrepancies between different versions of {\sc 21cmFAST} might also contribute in small part to this deviation: \citetalias{2020MNRAS.499.1640M} used {\sc 21cmFASTv1} \citep{2011MNRAS.411..955M}, which thus influences the range of template parameters, while the forecast in \citetalias{2021MNRAS.508.1262M} used {\sc 21cmFASTv2} \citep{2019MNRAS.484..933P}.}
(1) Our template ignores the dependence of the memory of reionization with the amplitude and tilt of the matter power spectrum. Both parameters play a key role in determining how many UV sources will be present in the epoch of reionization (\citetalias{2021MNRAS.508.1262M}). Ignoring this dependency leads to a decrease in the constraining power of cosmological information. (2) We marginalize each redshift bin separately, to compute the overall Fisher matrix. 
Specifically, we marginalize our Fisher matrix on a per redshift-bin basis before inverting the covariance and recovering the reduced Fisher matrix that involves only cosmological information for a given redshift bin. Then, we sum the reduced Fisher matrices for all redshift bins to obtain the full reduced Fisher matrix. 
We choose this strategy because the parameters of our template, which are nuisance parameters for the purpose of forecast of cosmological parameter constraints, contain the redshift evolution of $P_{m,\psi}$, and hence need to be treated as independent across different redshift bins, and marginalized on a per-redshift basis. Naturally, this strategy leads to a lesser forecast due to treating each parameter set by itself. (3) Uniform weighting for all reionization models obtained in \citetalias{2020MNRAS.499.1640M} (which were used to construct Table \ref{tab:temp_params}) may hamper our forecast, since quite early or late reionization models are not as likely as less extreme scenarios. We explore the improvement of not assigning equal probability to all reionization models in Appendix~\ref{app:priors} using informative priors. 


An interesting feature, which is slightly obscured in the forecasting contours, is that the performance for the amplitude of the matter power spectrum for the template is similar to that obtained using the high-resolution simulations of \citetalias{2021MNRAS.508.1262M}: $\sigma_{A_1}^{\rm Temp}/\sigma_{A_1}^{\rm MM21} = 1.25$. In contrast, the tilt of the matter power spectrum is not so well constrained using the template as the result of \citetalias{2021MNRAS.508.1262M}: $\sigma_{n_1}^{\rm Temp}/\sigma_{n_1}^{\rm MM21} = 1.69$. 
This reflects the fact that $A_1$ is slightly less degenerate with the template parameters $\alpha_{\rm re}$ and $\beta_{\rm re}$, and more degenerate with the template parameter $A_{\rm re}$, than $n_1$.

We highlight that the overall effect of the discrepancy between the template and the simulations at small scales (see Figure \ref{fig:perf}) is negligible in the forecast. This is because the derivatives with respect to the template parameters are at least an order of magnitude smaller at $k \geq 0.5$~Mpc$^{-1}$ than the ones at smaller wavenumber, e.g.\ $k = 0.14$ Mpc$^{-1}$. Note that the strength of the memory of reionization drops rapidly with increasing wavenumber (see, e.g., Figure 5 of \citetalias{2020MNRAS.499.1640M}). In addition, the derivatives with respect to $A_{\rm re}$ and $\beta_{\rm re}$ at small scales are usually smaller by an additional order of magnitude than their large-scale counterparts. Of course, this can translate into an even smaller effect of the small scales in the Fisher forecast due to the presence of two weighted derivatives in the matrix elements. Note, however, that if one would {\it compute} the 1D \lya power spectrum by integrating the 3D \lya power spectrum over $k_\perp$ using this template, the discrepancy at large $k$ would, unfortunately, be a difficult challenge to overcome because of the integration over $k_\perp$. For this reason, in Appendix~\ref{app:p1d}, we propose a template for the memory of reionization on the \lya forest 1D power spectrum similar to the 3D case, which we deem successful enough to reproduce our results using high-resolution simulations. 

Finally, we remind the reader that even though the simulations used in \citetalias{2021MNRAS.508.1262M} are more accurate than the analytical template proposed here, the former is susceptible to the details in the modeling of reionization \citep{2021MNRAS.504.1555M}, e.g., whether to use a constant ionization efficiency for photons to escape from their parent galaxies to the intergalactic medium versus a variable efficiency that depends on halo mass. In contrast, the template approach is not hindered by the modeling of cosmic reionization.

Throughout this work we have neglected the direct impact of \ion{He}{II} reionization \citep{2008ApJ...681....1F,2017ApJ...841...87L,2020MNRAS.496.4372U} since it is absent in both MM20 and MM21 simulations; however, we highlight that, at lower redshifts ($z \lessapprox$ 3), \ion{He}{II} reionization likely dominates over the memory of \ion{H}{I} reionization. Estimating the impact of \ion{He}{II} reionization on the \lya flux power spectrum is a demanding task and estimations vary from $\sim$ percentage to tens of percentage level disruption in the 1D \lya flux, but with agreement that the strongest effect is present at small scales ($k \gtrapprox 0.1 $ s/km) \citep{2017ApJ...841...87L,2020MNRAS.496.4372U}, which is the opposite trend of the \ion{H}{I} memory of reionization. In contrast, the impact on the 3D power spectrum is localized to large scales ($k \lessapprox 0.1 h/$Mpc$^{-1}$) and may result from significant correlations in the radiation field due to quasar emission \citep{2017ApJ...841...87L}. Naturally, a direct computation of the impact of \ion{He}{II} reionization on the memory of \ion{H}{I} reionization would require a self-consistent simulation capable of handling the difficulties of modeling cosmic helium II reionization, e.g. tackling both the long and short mean free path ionizing photons, and capable of resolving the HEMD phase of the temperature-density relation, simultaneously.

We leave this extensive exploration to future work. For now, we underscore that estimating the effect of \ion{He}{II} reionization in the memory of \ion{H}{I} reionization without a full implementation of patchy helium reionization in our small-scale simulations is a highly complex endeavor because of two competing effects: additional photoheating and less overall recombinations. Additional photoheating in \ion{He}{III} regions will lead to a faster relaxation of the HEMD gas. However, the injected energy would reduce the number of recombinations that the gas goes through. Overall, the result is likely a delay in the relaxation process, thus leading to a longer-lived HEMD phase and therefore stronger thermal relics (albeit associated with both \ion{H}{I} and \ion{He}{II} reionization). \citet{2018MNRAS.474.2173H} (see their \S6.3) showed that the addition of \emph{sudden} helium reionization results -- perhaps unsurprisingly -- in a decrease of the sensitivity to thermal relics from hydrogen reionization of approximately fifty percent at $z = 2.5$, specifically $\partial \ln T (z = 2.5) / \partial \ln T(z = 8) = 0.031$ (no sudden \ion{He}{II} reionization) compared to $\partial \ln T (z = 2.5) / \partial \ln T(z = 8) = 0.017$ (with sudden \ion{He}{II} reionization). Naively, this could imply that our results presented here, for the flux power spectra,  may be off by roughly a factor of two at low redshifts ($z < 3$).

\section{Summary} 
\label{sec:sum}
The DESI survey has already started and with it comes new challenges that will be tackled to assess the accuracy of the upcoming exciting new results. In particular, this work focuses on the impact of \ion{H}{I} reionization on the \lya forest. While this effect can be used to constrain the astrophysics of reionization (\citetalias{2021MNRAS.508.1262M}), this systematic is a challenge that can bias the inference of cosmological parameters from the \lya forest power spectrum. The main concern with this systematic is that simulations capable of tracking down the way the small-scale structure reionizes for cosmological time scales are computationally expensive \citep{2018MNRAS.474.2173H}. Thus the simple marginalization of reionization parameters with conventional MCMC methods in a Bayesian framework would prove out of reach with current computational capabilities. 

Here we propose an analytical template for the memory of hydrogen reionization in the \lya forest power spectrum, inspired by the Yukawa potential. We show that this Yukawa-like template is able to marginalize the effect of \ion{H}{I} reionization at small computational costs in a model-independent manner, with a slight overestimation of the errors for cosmological parameters. Specifically, we obtain the ratio of the forecast $1\sigma$ error $\sigma_{A_1}^{\rm Temp}/\sigma_{A_1}^{\rm MM21} = 1.25$ for $A_1$ (the amplitude of the linear matter power spectrum at $k_1 \equiv 2\pi h $ Mpc$^{-1}$), and $\sigma_{n_1}^{\rm Temp}/\sigma_{n_1}^{\rm MM21} = 1.69$ for its tilt $n_1$. 
The overestimation of the errors is likely due to three reasons: (1) neglecting the dependence of the memory of reionization on cosmology, (2) marginalizing the Fisher matrix on a per redshift-bin basis because the template parameters are independent across different redshift bins, and (3) adopting uniform weights of reionization histories. 

We also demonstrate that our Yukawa-like template is sufficient to reproduce the broadband effect of the memory of reionization in the \lya forest correlation function, and also able to robustly determine the expected bias of cosmological parameters due to the contamination sourced by the memory of reionization. 

Future work will be made to improve the marginalization results using this template. e.g.\ sampling the values of the template parameters that correspond to the allowed reionization histories in a more robust manner. In addition, future work will apply the template specialized for the 1D \lya power spectrum  -- introduced in Appendix \ref{app:p1d} -- to real data in order to obtain unbiased cosmological parameters with respect to the effect of inhomogeneous reionization in the \lya forest, and to place the first constraint on the global reionization history using the eBOSS data of \lya forest at $z \leq 4.0$. 

\section*{Acknowledgements}
We thank the anonymous referee for their insightful suggestions. We thank the DESI collaboration for providing the quasar luminosity function and spectrograph performance needed to make a realistic forecast. We are also grateful to Vid Ir\v{s}i\v{c}, Bohua Li, and Jiaxin Wu for useful comments and discussions.
This work is supported by National Key R\&D Program of China (Grant No.~2018YFA0404502), NSFC (Grant No.~11821303, 12050410236), and National SKA Program of China (Grant No.~2020SKA0110401). PMC was supported by the Tsinghua Shui Mu Scholarship. We acknowledge the Tsinghua Astrophysics High-Performance Computing platform at Tsinghua University for providing computational and data storage resources that have contributed to the research results reported within this paper.

\section*{Data Availability}
The data underlying this article will be shared on reasonable request to the corresponding authors.



\bibliographystyle{mnras}
\bibliography{template} 



\appendix
\section{Weights of reionization models}
\label{app:priors}
One of the main applications of the Yukawa-like template for $P_{m,\psi}$ is unbiased (with respect to the memory of reionization in the \lya forest) Bayesian inference of cosmological parameters with the \lya forest 3D power spectrum. In this context, Table \ref{tab:temp_params} gives the priors on the template parameters. However, using the flat priors on those parameters disregards an important, additional source of information --- current constraints on the timeline of cosmic reionization.

\begin{figure}
    \centering
    \includegraphics[width=\linewidth]{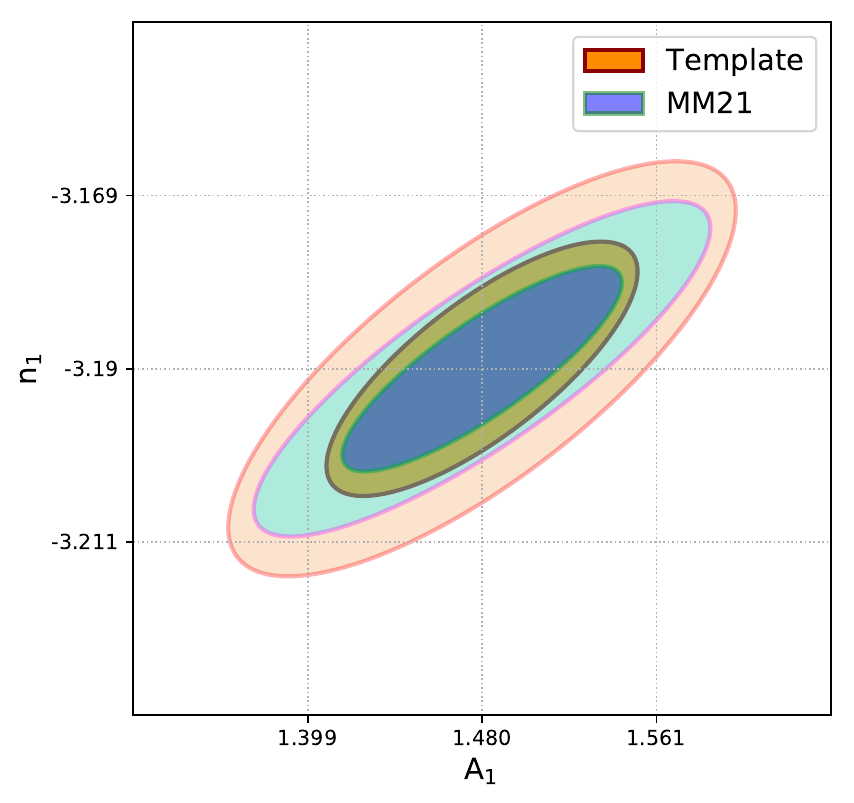}
    \caption{Same as Figure \ref{fig:forecast} but here we show the 1$\sigma$/2$\sigma$ contours (orange/light-orange ellipses) obtained using the Yukawa-like template using Gaussian priors based on the range of the template parameters in Table \ref{tab:temp_params}.}
    \label{fig:forecast_prior}
\end{figure}

In this section, we assume that the template parameters follow a normal distribution and assign informative priors based on the midpoint of reionization, which is currently constrained to likely occur at $z \approx 7.7$ \citep{2020A&A...641A...6P}. Specifically, the normal distribution is chosen in such a manner that the 1$\sigma$ range on each parameter disfavours both early and late reionization models, and it approximately coincides with the 1$\sigma$ range obtained by \citet{2020A&A...641A...6P}. We show the resulting forecast in Figure \ref{fig:forecast_prior}. The marginalization with the Gaussian priors recovers the morphology of forecast errors in cosmological parameter space more closely to the result of \citetalias{2021MNRAS.508.1262M} than using the flat priors, which is an encouraging sign of the impact of the weights of different reionization models in the template. (However, we caution that Table \ref{tab:temp_params} is constructed with a limited number of reionization models, which means that a more robust sampling of the parameter space is necessary to guarantee reliable results.) Likewise, future efforts in this direction will benefit greatly from near-term constraints on the timeline of cosmic reionization.  

\section{template Specialized for the memory of reionization in the 1D flux power spectrum}
\label{app:p1d}


\begin{figure*}
    \centering
    \includegraphics[width=\linewidth]{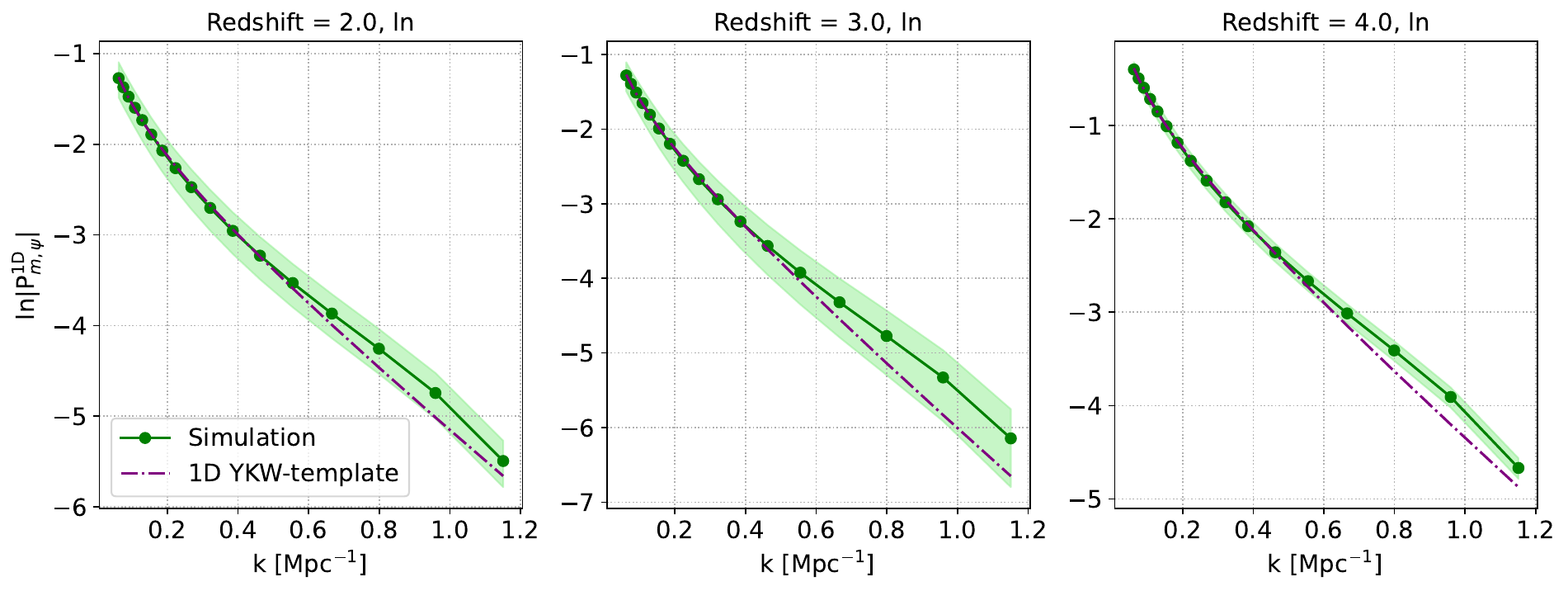}
    \caption{Performance of the Yukawa-like template for the memory of reionization imprinted in the {\it one-dimensional} \lya forest. Shown are the 1D cross-power spectra of the matter and transparency of the IGM, $P^{\rm 1D}_{m,\psi}(k,z_{\rm obs})$, in the natural logarithm at $z_{\rm obs} = 2.0$ (left), $3.0$ (middle) and $4.0$ (right), respectively. We show the results from the simulations of \citetalias{2020MNRAS.499.1640M} (green dots) as the fiducial model, with the light-green shaded area covering the error computed in \citetalias{2020MNRAS.499.1640M}. In comparison, we show the 1D Yukawa-like template (purple dash-dotted line) that is the best-fit to the simulation results.}
    \label{fig:perf-1d}
\end{figure*}

As the complement to the 3D broadband measurements, the \lya forest 1D power spectrum requires the cross-correlation between pixels along the same line of sight. Mathematically, the expression is given by integrating the 3D counterpart over the perpendicular direction (i.e. $k_{\perp}$) to the line of sight \citep{2013A&A...559A..85P}, taking the form (see Eq.~10 of \citealt{2019MNRAS.487.1047M} for a detailed derivation)
\begin{eqnarray} \label{eq:p1d}
    P^{1\mathrm{D}}_F(z,k)=b^2_F(z)P^{1\mathrm{D}}_m(z,k)+2b_F(z)b_{\Gamma}(z)P^{1\mathrm{D}}_{m,\psi}(z,k) \, ,
\end{eqnarray}
where $b_F$ and $b_{\Gamma}$ are the bias parameters that relate flux fluctuations to density and radiation fluctuations, respectively. The first term of Eq.~(\ref{eq:p1d}) is the conventional 1D \lya forest power spectrum, which includes the non-linear effects attributed to the non-linear growth, peculiar velocities and pressure and thermal broadening \citep{2003ApJ...585...34M, 2015JCAP...12..017A}. In contrast, the second term corresponds to the memory of reionization in the 1D \lya forest power spectrum. Hence, it contains the 1D analog of the cross-power spectrum of matter and transparency of the IGM (see Eq.~\ref{eq:psi}), which parametrizes how transparent a given patch of the sky that reionizes at a given redshift is compared to another patch that reionizes at a different redshift, and cross-correlates it with the distribution of matter to capture the inside-out nature of cosmic reionization.

Regarding the 3D-to-1D mapping, parametrization for the cross-power spectrum of matter and transmission $P^{1\mathrm{D}}_{m,\psi}$ is motivated by integrating Eq.~(\ref{eq:psi}) over $k_\perp$. We find that the analytic form of the template Eq.~(\ref{eq:yukawa}) remains unchanged for the 1D \lya forest power spectrum. The Yukawa-like template also preserves the physical meaning of the three template parameters, only that the variation of $\alpha_{\rm re}$ with different reionization scenarios is affected by the integration. While early reionization models generally mean small amplitudes $A_{\rm re}$ and large exponents $\beta_{\rm re}$ in consistency with the 3D template, large values of $\alpha_{\rm re}$ are favoured in the 1D template, which is in an opposite manner as the 3D template behaves. We show the performance of the 1D template with respect to the fiducial model of \citetalias{2020MNRAS.499.1640M} in Figure \ref{fig:perf-1d}. The predicted results by our 1D template begin to depart from the high-resolution simulations of \citetalias{2020MNRAS.499.1640M} at $k = 0.4$~Mpc$^{-1}$, which is similar to the 3D case.

Meanwhile, simply applying this analytic form to the 1D \lya flux power would lead to much less stringent constraints on cosmology than what we intend to extract. This is primarily owing to (1) the weak impact of inhomogeneous reionization on the 1D power spectrum (e.g., its strength at $k = 0.14$~Mpc$^{-1}$ stands only at $\sim$ 15 \% at $z = 4.0$ --- see Fig.~\ref{fig:1D effect} and \citetalias{2020MNRAS.499.1640M}), and (2) the integration of the 3D power spectrum exaggerating the small-scale discrepancy (see Fig.~\ref{fig:perf}). 

Concerning the fact that the original template neglects the role of cosmology, we design a proxy that allows for the dependence of the analytic template on the cosmological parameters $A_1$ and $n_1$. The specialized template also partially takes such dependence of template parameters into account by assigning informative priors based on a range of cosmological models. The reionization histories from \citetalias{2020MNRAS.499.1640M}, together with these models, inform the fashion in which the total power varies with reionization and cosmological parameters. While the former takes the memory of reionization into account in the same way as the 3D analysis, the latter aids in the partial involvement of cosmology in the \lya forest 1D power spectrum.
Table~\ref{tab:cosmo_model} lists all five models that we apply to the Fisher analysis. Each cosmological model corresponds to a combination of different values of the two cosmological parameters, yielding one fiducial and four other cosmological models. The parameter range for the Yukawa-like 1D template is given in Table~\ref{tab:temp_params_1d} at the lowest and highest redshift bins of the 1D survey.

\begin{table}
    \centering
    \caption{Cosmological models with different combinations of the r.m.s.\ mass fluctuation $\sigma_8$ and the spectral index $n_s$. The two parameters are varied in the fashion that one is kept fiducial and the other one is adjusted. The symbol "---" means the parameter here takes the fiducial value.}
    \label{tab:cosmo_model}
	\begin{tabular}{ccc}
		\hline\hline
		Cosmology & $\sigma_8$ & $n_s$ \\
		\hline
		fiducial & 0.8159 & 0.9667 \\
		\hline
		$\sigma_8$+ & 0.8920 & --- \\
		$\sigma_8$- & 0.7325 & --- \\
		$n_s$+ & --- & 1.04 \\
		$n_s$- & --- & 0.895 \\
		\hline\hline
    \end{tabular}
\end{table}

\begin{table}
    \centering
    \caption{Parameter range for the 1D Yukawa-like template for the high redshift bins. Values are obtained by fitting to the cosmological models listed in Table~\ref{tab:cosmo_model} and to the reionization models presented in \citetalias{2020MNRAS.499.1640M}.}
    
    \begin{tabular}{ccc}
        \hline\hline
         Template parameter & $z = 3$ & $z = 4$ \\
         \hline
         $A_{\rm re}$ & [0.0768, 0.1921] & [0.2235, 0.3984]\\
         $\alpha_{\rm re}$ & [3.1724, 5.1430] & [2.8907, 3.1913] \\
         $\beta_{\rm re}$ & [0.2954, 0.4975] & [0.2799, 0.4615] \\
         \hline\hline
    \end{tabular}
    \label{tab:temp_params_1d}
\end{table}

\begin{figure}
    \centering
    \includegraphics[width=\linewidth]{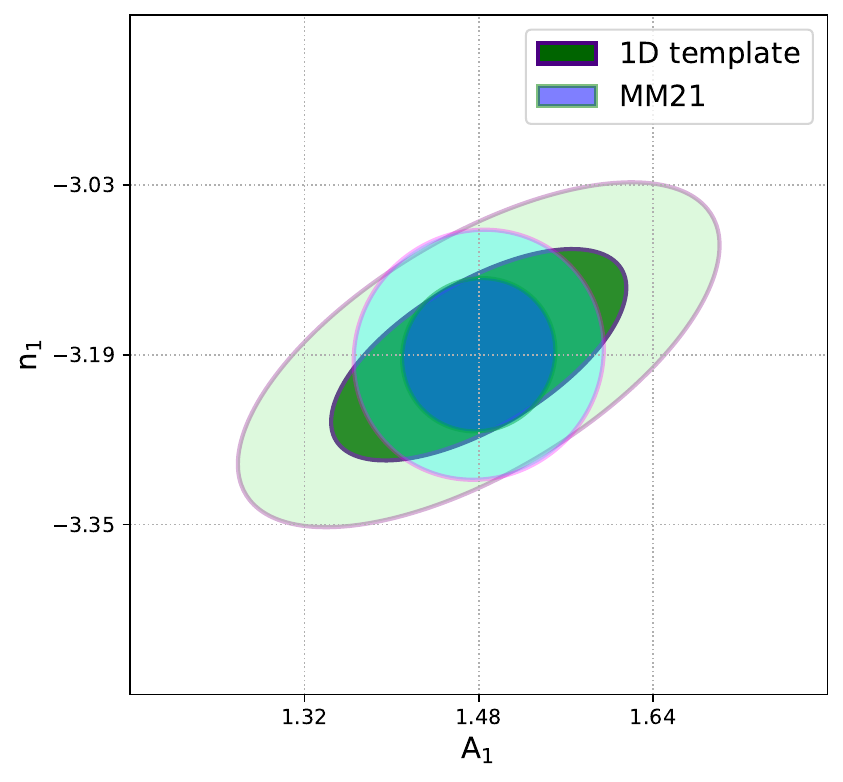}
    \caption{Forecast for extraction of cosmology from the 1D \lya forest power spectrum at the redshift range 3.0 $\leq z \leq$ 4.0. Here we implement the specialized template using informative priors obtained from the non-fiducial cosmological models listed in Table~\ref{tab:cosmo_model} to partially account for the cosmological models with non-fiducial values of $\sigma_8$ and $n_s$. We show the 1$\sigma$/2$\sigma$ contours obtained by \citetalias{2021MNRAS.508.1262M} using high-resolution simulations (blue/cyan ellipses), and the 1$\sigma$/2$\sigma$ contours obtained using the 1D Yukawa-like template (green/light-green ellipses), respectively. }
    \label{fig:forecast_1D_prior}
\end{figure}

\begin{figure}
    \centering
    \includegraphics[width=\linewidth]{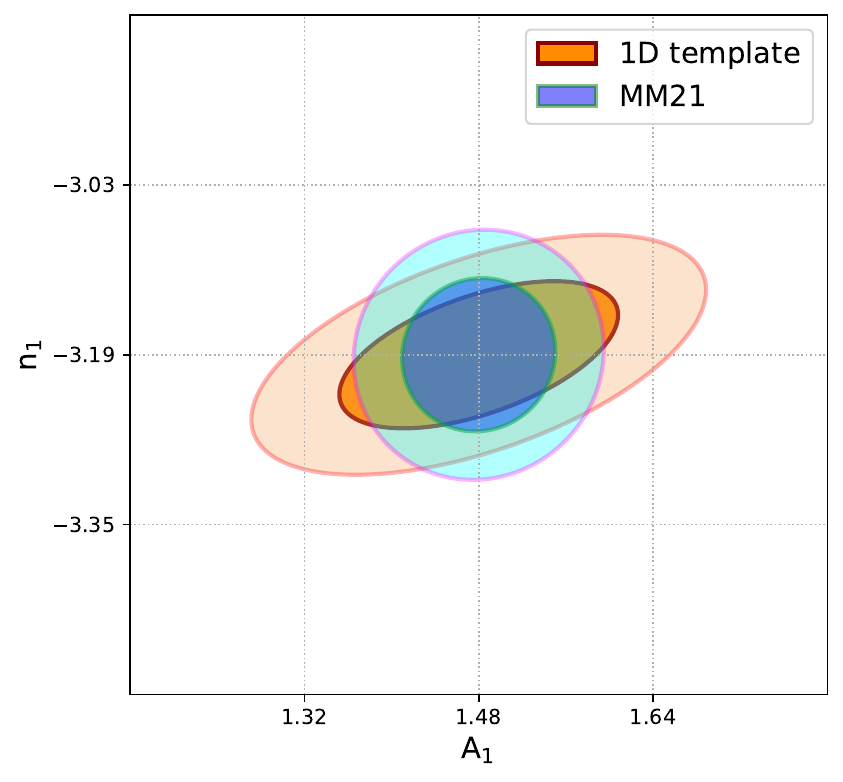}
    \caption{Same as Figure~\ref{fig:forecast_1D_prior} but here we show the 1$\sigma$/2$\sigma$ contours (orange/light-orange ellipses) obtained using the 1D Yukawa-like template by adding priors from HERA to disfavour uncommonly early and late reionization scenarios. These extreme conditions are selected based on the reionization models in Table 3 of \citet{2017MNRAS.472.2651G}. }
    \label{fig:forecast_1D_hera}
\end{figure}

The Fisher forecast for the 1D template is performed on the same basis as that described in \S\ref{sec:forecast} but here for an ideal eBOSS-like survey (see Section 3 of \citetalias{2021MNRAS.508.1262M} for a detailed description), i.e. we use Eq.~(\ref{eq:fisher}) with no $k$-cutoff, the parameter vector is now $p_i = \{\bar{F}, A_1, n_1, A_{\rm re}, \alpha_{\rm re}, \beta_{\rm re} \}$, $\sigma_{z_i, k_j}$ is the eBOSS uncertainty for the corresponding bin (here taken from \citealt{2019JCAP...07..017C}), and the flux power, which includes the memory of reionization, is given by Eq.~(\ref{eq:p1d}). 

Besides, the forecast is performed for the redshift bins 3.0 $\leq$ $z$ $\leq$ 4.0 due to the more evident memory of \ion{H}{I} reionization. This redshift range can also shield the HEMD phase of the temperature-density relation \citep{2018MNRAS.474.2173H} from the summit of \ion{He}{II} reionization when the thermal state of the IGM has not yet been fully impacted. For each non-fiducial model in Table~\ref{tab:cosmo_model}, we compute a reduced Fisher matrix that encodes the uncertainty information for the transmitted flux and the three template parameters. The informative prior is combined via  
\begin{eqnarray}
    \sigma_{p^{(i)}}^{2} = \sigma_{\sigma_8,p^{(i)}}^{2} + \sigma_{n_{s},p^{(i)}}^{2}\,, \label{eq:cosmo_prior}
\end{eqnarray}
Within the same type of cosmology, the propagation of uncertainty is the summation of fractional uncertainty 
\begin{eqnarray}
    \sigma_j^2 = \left[\left(\frac{\sigma_{j+}}{ p^{(i)}_{j+}}\right)^2 + \left(\frac{\sigma_{j-}}{p^{(i)}_{j-}}\right)^2\right]\cdot (p^{(i)}_{\mathrm{fid}})^2\,, \label{eq:cosmo_sigma}
\end{eqnarray}
where $p^{(i)}$ refers to the template parameters $A_{\rm re}$, $\alpha_{\rm re}$ and $\beta_{\rm re}$, and $j = \{\sigma_8\,, n_{s}\}$. Eq.~(\ref{eq:cosmo_sigma}) includes a normalization scheme that the error from each model is scaled by a factor of $(p^{(i)}_{\mathrm{fid}}/p^{(i)}_j)^2$. Intuitively, a direct consequence of allowing for more possibilities is a reduction in precision.  This implication is revealed physically from Eq.~(\ref{eq:cosmo_prior}) and Eq.~(\ref{eq:cosmo_sigma}) that the limiting case of reckoning with more cosmological models is to yield a larger prior and hence a smaller Fisher matrix element for a given template parameter, and thus a larger error.

We show the ability of our 1D template to extract cosmology from the 1D \lya forest power spectrum in Figure~\ref{fig:forecast_1D_prior}. The implementation of the template leads to a particular overestimation of $A_1$ and also a slight change in the direction of degeneracy. This is quantified in terms of the ratios $\sigma_{A_1}^{\mathrm{Temp},\mathrm{1D}}/\sigma_{A_1}^{\mathrm{MM21,\mathrm{1D}}} = 1.93$ and $\sigma_{n_1}^{\mathrm{Temp},\mathrm{1D}}/\sigma_{n_1}^{\mathrm{MM21},\mathrm{1D}} = 1.38$, where the tilt of the linear matter power spectrum is more constrained. This is just opposite to the forecast by the 3D template where the amplitude is better forecasted (see Section~\ref{sec:sum}), because a large proportion of the total 1D \lya forest power is accounted for by the linear matter spectrum (see also Eq.~\ref{eq:p1d}) that is more sensitive to the spectral index than to the amplitude.

In addition to cosmological priors, which inform us about reasonable cosmological models,  we also improve the template by considering constraints from other probes. However, using the same Gaussian priors obtained from \citet{2020A&A...641A...6P} as in the 3D analysis (see Appendix~\ref{app:priors}) will result in an over-constrained $A_1-n_1$ ellipse and hence we take the opportunity to improve on the informative priors. This is primarily due to the insensitivity of the 1D power spectrum to different reionization scenarios that are discussed before and also the overestimation of the 1$\sigma$ range on each parameter. The latter is a direct result of using the \citetalias{2020MNRAS.499.1640M} models that correspond to the points on the $T_{\mathrm{min}}-\zeta$ axes (see Table 1 of \citetalias{2020MNRAS.499.1640M}) and therefore have neglected the off-axis regime that also satisfies the Planck constraints. Nevertheless, this is inevitable because a comprehensive translation of the neutral fraction constraints into constraints on $T_{\mathrm{min}}$ and $\zeta$ is not based on a one-to-one mapping and therefore is not feasible.\footnote{This mapping can be easily implemented in a recent semi-numerical code for reionization modelling {\sc amber} \citep{2021arXiv210910375T} from ionization fraction $\bar{x}_i(z)$ to the midpoint of reionization $z_{\mathrm{mid}}$, duration $\Delta z$ and asymmetry $A_z$. However, the reionization parameters we adopt in this work are derived from the excursion set formalism and are therefore not applicable to the mapping.} Regarding a more direct constraint on reionization parameters, we have applied the priors from a mock 1000h observations with HERA \citep{2017MNRAS.472.2651G} to assign a lesser weight to uncommonly early and late reionization scenarios. Figure~\ref{fig:forecast_1D_hera} exhibits this creditable reduction in errors on the amplitude and tilt by 8.9\% and 44\% respectively. The remarkable improvement of the latter is attributed to the sensitivity of the major component of the $P^{1\mathrm{D}}_F$ (i.e. $P^{1\mathrm{D}}_m$) to the spectral index.
This improvement is also reflected from the comparisons with high-resolution simulations that $\sigma_{A_1}^{\mathrm{HERA},\mathrm{1D}}/\sigma_{A_1}^{\mathrm{MM21,\mathrm{1D}}} = 1.82$ and $\sigma_{n_1}^{\mathrm{HERA},\mathrm{1D}}/\sigma_{n_1}^{\mathrm{MM21},\mathrm{1D}} = 0.96$. These preliminary results are also a sign that the precision of $n_s$ we forecast with the template can be comparable to that obtained with simulations if a joint analysis with HERA is performed. Moreover, as HERA is already operational and will likely release their major results on a similar timeline to that of DESI, our first attempt at an analytical template for the memory of reionization in the 1D \lya flux power spectrum can be used as a proxy for the joint analysis that is likely to be performed between these two probes in the incoming future.

Finally, we summarize the sources of the overestimation of errors compared to simulations as (1) the per redshift-bin basis and (2) the potential limitations of factoring in cosmology using the proxy we proposed above. The former naturally takes the assumption that measurements at different redshift bins are mutually independent. However, the parameters applied to the forecast will evolve with redshift, which leads to an inconsistency that overestimates the uncertainty of the analysis. Furthermore, we also raise concerns on the 1D analysis with high redshift bins. These redshifts, despite saving the \lya forest power from the whole effects of \ion{He}{II} reionization, could bring a dearth of reliability for the forecast due to our empirical redshift evolution that adopts a pivot redshift at $z = 2.25$ based on the simulated data presented in \citet{2003ApJ...585...34M}. However, we deem our choice of redshift bins lying well in an acceptable range from the pivot that is later extrapolated to $z = 3.0$ by \citet{2015JCAP...12..017A}. Besides, the focus here is to evaluate the performance of our specialized template after including cosmology, which we reckon practicable at this level to partially account for the memory of reionization without expensive simulations. Hence we leave implementations of \ion{He}{II} reionization and further appraisal of high redshift evolution to future work. Moreover, we would like to emphasize here that our 1D template is still at a primitive stage for analytic parametrization. Related work will continue to explore the possibility of reflecting cosmology in a more straightforward manner and of applying the ameliorated template to real observation data from eBOSS and DESI.


\bsp	
\label{lastpage}
\end{document}